\documentclass[twocolumn,times]{aastex62}
\usepackage{apjfonts}

\newcommand{\nthp}{$\mathrm{N_2 H^+}$}
\newcommand{\ctfs}{$\mathrm{C^{34}S}$}
\newcommand{\hmol}{$\mathrm{H_2}$}
\newcommand{\kmps}{$\mathrm{km\;s^{-1}}$}

\received{September 11, 2019}
\revised{February 12, 2020}
\accepted{February 14, 2020}
\submitjournal{ApJ}

\shorttitle{CS Depletion in Prestellar Cores}
\shortauthors{S. Kim et al.}

\begin{document}

\title{CS Depletion in Prestellar Cores}
\correspondingauthor{Chang Won Lee}
\email{cwl@kasi.re.kr}

\author[0000-0001-9333-5608]{Shinyoung Kim}
\affiliation{Korea Astronomy and Space science Institute, 776 Daedeok-daero, Yuseong-gu, Daejeon 34055, Republic of Korea}
\affiliation{University of Science and Technology, Korea, 217 Gajeong-ro, Yuseong-gu, Daejeon 34113, Republic of Korea}

\author{Chang Won Lee}
\affiliation{Korea Astronomy and Space science Institute, 776 Daedeok-daero, Yuseong-gu, Daejeon 34055, Republic of Korea}
\affiliation{University of Science and Technology, Korea, 217 Gajeong-ro, Yuseong-gu, Daejeon 34113, Republic of Korea}

\author{Maheswar Gopinathan}
\affiliation{Indian Institute of Astrophysics, 2nd Block, Koramangala, Bengaluru, Karnataka 560034, India}

\author[0000-0002-2569-1253]{Mario Tafalla}
\affiliation{IGN, Observatorio Astron\'{o}mico Nacional, Calle Alfonso XII, E-28014 Madrid, Spain}

\author{Jungjoo Sohn}
\affiliation{Korea National University of Education, 250 Taeseongtabyeon-ro, Gangnae-myeon, Heungdeok-gu, Cheongju-si, Chungbuk 28173, Republic of Korea}

\author{Gwanjeong Kim}
\affiliation{National Astronomical Observatory of Japan, 462-2 Nobeyama, Minamimaki, Minamisaku, Nagano 384-1305, Japan}

\author{Mi-Ryang Kim}
\affiliation{Korea Astronomy and Space science Institute, 776 Daedeok-daero, Yuseong-gu, Daejeon 34055, Republic of Korea}

\author[0000-0002-6386-2906]{Archana Soam}
\affiliation{SOFIA Science Center, USRA, NASA Ames Research Center, MS-12, N232, Moffett Field, CA 94035, USA}

\author[0000-0002-2885-1806]{Philip C. Myers}
\affiliation{Center for Astrophysics, Harvard \& Smithsonian, 60 Garden Street, Cambridge, MA 02138, USA}

\begin{abstract}

The CS molecule is known to be adsorbed onto dust in the cold and dense conditions, causing it to get significantly depleted in the central region of cores.
This study is aimed to investigate the depletion of the CS molecule using the optically thin \ctfs\ molecular line observations. 
We mapped five prestellar cores, L1544, L1552, L1689B, L694-2 and L1197 using two molecular lines, \ctfs\ $(J=2-1)$ and \nthp\ $(J=1-0)$ with the NRO 45-m telescope, doubling the number of cores where the CS depletion was probed using \ctfs. 
In most of our targets, the distribution of \ctfs\ emission shows features that suggest that the CS molecule is generally depleted in the center of the prestellar cores. 
The radial profile of the CS abundance with respect to \hmol\ directly measured from the CS emission and the \textit{Herschel} dust emission indicates that the CS molecule is depleted by a factor of $\sim 3$ toward the central regions of the cores with respect to their outer regions. 
The degree of the depletion is found to be even more enhanced by an order of magnitude when the contaminating effect introduced by the presence of CS molecules in the surrounding envelope that lie along the line-of-sight is removed. 
Except for L1197 which is classified as relatively the least evolved core in our targets based on its observed physical parameters, we found that the remaining four prestellar cores are suffering from significant CS depletion at their central region regardless of the relative difference in their evolutionary status.

\end{abstract}


\section{Introduction} \label{sec:int}

The formation of low mass stars is understood based on a sequence of conceptually different stages \citep{Larson:1969cq,Shu:1987dp}. 
The first stage in this sequence corresponds to the fragmentation of molecular clouds into a number of gravitationally bound cores or condensations that are initially supported against gravity by a combination of thermal, magnetic and turbulent pressures \citep{Mouschovias:1991kd,Shu:1987dp}. 
When these condensations become dense enough to be detected with high density tracers such as, \nthp\ or CS transitional lines, but have not yet formed any central hydrostatic object, they are called \textit{starless} cores \citep{Benson:1983go,Lee:1999fi}. 
Some of these starless cores that are gravitationally bound, show extended infall asymmetry and are on the verge of star formation, are called \textit{prestellar} cores \citep{WardThompson:2007tx}. 
These cores have a typical size of 0.1 pc and a few solar masses of gas and dust \citep{Kirk:2007hi}. 
Thus prestellar cores are ideal objects to study the physics of core contraction and hence the initial phases of star formation.

The CS and \nthp\ rotational transition lines being relatively optically thick and thin, respectively, have been widely used in the study of the core contraction. 
The blue asymmetric line profile in the CS line and the Gaussian profile of the optically thin \nthp\ line are widely used to trace infall motions in the prestellar cores. 
To conduct a detailed study of the infall motions, a systematic survey was carried out by \citet{Lee:1999de} on 220 starless cores using the CS and \nthp\ lines. 
They found that the inward contraction is statistically significant in their sample of starless cores. 
On the other hand, \citet{Leger:1983wr} and \citet{Bergin:1997fl} have shown that in the cold cores having a number density greater than a few $10^4$ cm$^{-3}$, molecule such as CS, which is relatively more polar, efficiently gets adsorbed onto the dust grains and depletes out, while non-polar molecule such as $\mathrm{N_2}$, because of its high volatile property, remains in the gaseous phase up to a density of a few $10^6$ cm$^{-3}$. 
\citet{Tafalla:2002bn}, based on mapping observations of five starless cores in CS $(2-1)$ line, showed that a strong CS depletion was found at the peak positions of dust continuum in all their five targets. 
Thus the depletion of CS molecules seems to be an important problem which can not be easily ignored in the core contraction studies. 
For example, if most of the CS molecules are removed from the core center, it may not be plausible to interpret the infall asymmetry observed at the peak positions as due to the inward motion of the material at the innermost parts of the core. 

Though the issue of the depletion of the CS molecules in the central region of the prestellar cores is important, the efforts to seek its observational evidence have been rather limited. 
Because the CS $(2-1)$ line is optically thick in the high-density central region of the prestellar cores \citep{Lee:1999de}, using this line to study the distribution of CS in the cores can be hampered due to its saturation. 
Instead, mapping observations using its rare isotopologue molecule, \ctfs, line may allow us to make a direct estimation of the distribution of CS molecules in the central region of the dense cores \citep[e.g.,][]{Tafalla:2002bn,Tafalla:2004jx}. 
So far only a few observational attempts have been made to characterize the CS depletion in the central region of the dense cores ($N_\mathrm{H_2} \gtrsim 10^{22}$ cm$^{-2}$) using \ctfs\ molecular line. 
The mapping study of only four starless cores have been carried out using \ctfs\ line as yet with sufficiently high spatial resolution and wide spatial coverage (L1521E: \citealt{Tafalla:2004ex}; L1498: \citealt{Tafalla:2004jx}; L492: \citealt{Hirota:2006fh}; MCLD123.5+24.9: \citealt{Heithausen:2008ji}). 
However, a detailed study on CS depletion was conducted only for L1498 \citep{Tafalla:2004jx}. 
A pointing survey with \ctfs\ was carried out for 23 starless cores belonging to the Chamaeleon I region using \ctfs\ molecular line \citep{Tsitali:2015co}. 
It was found that the \ctfs\ abundance decreases with the increase in the \hmol\ column density. 

In this study, we augment the sample of cores observed using \ctfs\ molecular line by presenting the results of high resolution ($<20\arcsec$) mapping observations of five prestellar cores to better characterize the CS depletion in them. 

In Sections \ref{sec:obs} and \ref{sec:res}, we describe the line and target selections for our observations and various analysis on CS distribution and their results. In Sections \ref{sec:disc} and \ref{sec:conc}, we present the discussions and conclusions for our results.

\section{Observations} \label{sec:obs}

\subsection{Line Selection} \label{sec:obs-line}

We used two molecular line tracers, \ctfs\ $(2-1)$ and \nthp\ $(1-0)$, to observe the five prestellar cores studied here. 
The \ctfs\ $(2-1)$ was chosen as the key molecular line in our study to investigate the distribution of CS molecule in dense cores instead of the main isotopologue, CS $(2-1)$ line because of its possible saturation or high optical depth especially for dense regions of the cores. 
Such saturation effect due to the high optical depth in the CS $(2-1)$ line was frequently seen in a form of double-peaked or skewed features in its profiles, indicating that the CS $(2-1)$ line is mostly optically thick toward dense cores \citep{Lee:1999de, Lee:2001jm}. 
On the other hand, \ctfs\ $(2-1)$ line is expected to be optically thin as the $\mathrm{C^{32}S}/\mathrm{C^{34}S}$ isotopic ratio in the interstellar medium (ISM) is found to be almost equivalent to its solar value of 22.7 \citep{Lucas:1998ue}. 
In L1544, \citet{Tafalla:1998gf} found that the \ctfs\ $(2-1)$ line was the only tracer that showed a single Gaussian shape without any significant saturation in the line profile, while all other tracers showed self-absorbed features. 
Whether the CS and \ctfs\ $(2-1)$ lines are optically thick or thin can be quantitatively inferred from our present NRO 45-m observation for L1552 with together our complementary CS $(2-1)$ data. 
The intensities of both these lines were found to be almost same at the central velocity towards the central region of the core and the measured $\tau_{\mathrm{CS}}$ was found to be about 22, implying that the CS $(2-1)$ line is highly optically thick while the maximum optical depth of \ctfs\ $(2-1)$ line is at most as large as about 1. 
Therefore the \ctfs\ $(2-1)$ line is expected to be fairly optically thin towards most of the central high-density regions of the prestellar cores and is thought to be an adequate tracer for investigating the real distribution of the CS molecules inside the cores studied here.

The second tracer used in this study is the \nthp\ $(1-0)$ molecular line. 
It is well known that compared to the CS molecule, the distribution of the \nthp\ $(1-0)$ line is chemically less affected even towards the very late stage of core evolution \citep{Bergin:2007iy, Crapsi:2005kp}. 
Therefore, this line is considered to be very useful to examine the evolutionary status of the prestellar cores having densities $>10^5\;\mathrm{cm^{-3}}$ and to compare with the distribution of the CS molecules inside the cores.

As for the line frequencies for two lines, we used 96.412953 GHz for \ctfs\ $(2-1)$ \citep{Gottlieb:2003gt} and 93.176258 GHz for \nthp\ $(JF_1F=101-012)$ \citep{Lee:2001jm} in our observations.

\subsection{Target Selection} \label{sec:obs-target}

\begin{deluxetable}{lccDc}
    \tablecaption{Observing Targets \label{tbl:sample}}
    \tablehead{
    \colhead{Core} & \colhead{R.A.\tablenotemark{a}} & \colhead{Dec.\tablenotemark{a}} & \multicolumn{2}{c}{$V_{\mathrm{N_2 H^+}}$\tablenotemark{b}} & \colhead{Distance\tablenotemark{c}} \\
    \colhead{} & \colhead{(J2000.0)} & \colhead{(J2000.0)} & \multicolumn{2}{c}{[$\mathrm{km \; s^{-1}}$]} & \colhead{[pc]}}
    \decimals
    \startdata
    L1544  & 05 04 16.0 & +25 11 03 &  7.09 & 140 \\
    L1552  & 05 17 39.0 & +26 05 00 &  7.65 & 140 \\
    L1689B & 16 34 48.6 & -24 38 03 &  3.53 & 165 \\
    L694-2 & 19 41 05.0 & +10 56 49 &  9.54 & 250 \\
    L1197  & 22 37 02.6 & +58 57 29 & -3.19 & 400 \\
    \enddata
    \tablenotetext{a}{Peak position of {\textit{Herschel}} 250 \micron\ continuum data}
    \tablenotetext{b}{From the hyperfine fitting for \nthp\ lines in this study}
    \tablenotetext{c}{\citet{Jijina:1999bg}}
\end{deluxetable}

The degree of CS depletion was known to increase in later stages of the evolving core and become significant in the collapsing phase \citep{Bergin:1997fl}.
Thus the targets for our observations were selected based on the presence of any evidence indicating that the cores are already at a highly evolved stage \citep{Lee:1999de,Lee:2001jm,Crapsi:2005kp}. 
The targets thus selected are namely, L1544, L1552, L1689B, L694-2, and L1197. 
All these cores are found to show infall asymmetric profiles in CS $(2-1)$ line. 
This implies that they are possibly in a state of gravitational contraction and soon become dense enough to initiate formation of (sub)-stellar object(s) \citep{Lee:2001jm}. 
All our target cores are considered as starless because of the absence of any IRAS/Spitzer/Herschel point sources \citep[e.g.,][]{Lee:1999fi}. 
They all are found to be dense (between $10^4$ and $10^5\;\mathrm{cm^{-3}}$), compact (optical size of $0.05-0.10$ pc), and have narrow line widths ($\Delta V_{\mathrm{FWHM}}$ of \nthp\ $\sim0.21-0.32$ \kmps) \citep{Lee:2001jm}. 
The targets and their basic information are given in Table \ref{tbl:sample}

\subsection{NRO-BEARS: Line Observations} \label{sec:obs-nro}

\begin{deluxetable}{lcc}
    \tablecaption{Summary of the Observational Parameters\label{tbl:obs}}
    \tablehead{
    \multicolumn{1}{l}{Telescope} & \multicolumn{2}{c}{NRO 45 m} \\
    \multicolumn{1}{l}{Date} & \multicolumn{2}{c}{2004 Feb 11-14} \\
    \multicolumn{1}{l}{Transitions} & \colhead{$\mathrm{C^{34}S}\;(2-1)$} & \colhead{$\mathrm{N_2 H^+}\;(1-0)$}}
    \startdata
    Rest frequency [GHz]             & 96.412953\tablenotemark{a} & 93.176258\tablenotemark{b} \\
    HPBW [$\arcsec$]            & 17.3                       & 17.9 \\
    Receiver                    & \multicolumn{2}{c}{BEARS; ($5\times5$) beams} \\
    Mapping mode                & \multicolumn{2}{c}{$4\times4$ pointing} \\
    Mapping area [$\arcsec$]    & \multicolumn{2}{c}{$226\times 226$} \\ 
    Grid separation [$\arcsec$] & \multicolumn{2}{c}{$20.55$} \\
    Spectrometer                & \multicolumn{2}{c}{Auto Correlator} \\
    Band width                  & \multicolumn{2}{c}{16 MHz (1024 channels)} \\ 
    Velocity resolution [\kmps] & 0.049 & 0.050 \\
    \enddata
    \tablenotetext{a}{\cite{Gottlieb:2003gt}}
    \tablenotetext{b}{\cite{Lee:2001jm}}
\end{deluxetable}

We observed our target cores in \ctfs\ $(2-1)$ and \nthp\ $(1-0)$ using the 45 m telescope of the Nobeyama Radio Observatory (NRO) during February 2004. 
In our observations the SIS 25-BEam Array Receiver System (BEARS) was used to make maps of an area of $226\arcsec \times 226\arcsec$ with $20\farcs 55$ separation for each source through $4 \times 4$ pointings. 
The sky was subtracted by observations in frequency switching mode. 
The backend that we used was an auto-correlator which was set to 16 MHz bandwidth mode of frequency resolution of 15.63 kHz corresponding to $\sim 0.05$ \kmps\ at the frequencies of \ctfs\ $(2-1)$ and \nthp\ $(1-0)$. 
The telescope beam size (HPBW) at the observing frequencies is approximately $17\farcs 3$ for \ctfs\ $(2-1)$ and $17\farcs 9$ for \nthp\ $(1-0)$. 
The main beam efficiency at observed wavelengths is $\sim 0.5$\footnote{\url{https://www.nro.nao.ac.jp/~nro45mrt/html/prop/eff/eff-intp.html}}. 

We note that 25 receivers in BEARS had different gains and thus their gains were needed to be calibrated in each. 
For this purpose we observed a bright position ($\mathrm{05^h32^m49^s.8}$ $-05\arcdeg 21\arcmin 23\farcs9$) of Orion-KL4 as a calibration source using two receiver systems, the BEARS and S100/S80 receiver. 
We observed the position of the calibration source with each beam of BEARS and then the same position with a single beam S100/S80 receiver which would give a reference intensity scale of the calibration source. 
The correction factors were calculated for individual horns of the BEARS by comparing the relative intensity obtained through each horn with that through the S100/S80 receiver. 
The correction factors were $0.95-1.73$ for the \nthp\ $(1-0)$ and $0.75-1.66$ for \ctfs\ $(2-1)$.
Our CS (2-1) line data for L1552 that were used for the complementary purpose of showing its high optical depth in this paper had been obtained with NRO 45m in March 2000 by using S100/S80 receiver and FX correlator.
The observational data were initially reduced using the NEWSTAR\footnote{\url{https://www.nro.nao.ac.jp/~nro45mrt/html/obs/newstar/}}, and then converted to CLASS format for further detailed reduction using the CLASS package of GILDAS\footnote{\url{http://www.iram.fr/IRAMFR/GILDAS}} software.

\subsection{\textit{Herschel} Continuum Data} \label{sec:obs-psw}

We used the \textit{Herschel} dust continuum data \citep{Andre:2010ka} as a proxy to the distribution of \hmol\ molecules in order to compare the distribution of the CS or \nthp\ with respect to the distribution of \hmol\ molecules. 
Because the dust continuum emission is not affected by any chemical changes, it can provide a true feature of the distribution of the \hmol\ molecules. The 160, 250, 350, and 500 $\micron$ images of our target cores were obtained from the \textit{Herschel} archive. 
The spatial resolutions corresponding to the full width half maximum (FWHM) of the {\textit{Herschel}}'s beam are $10\farcs7$, $17\farcs9$, $24\farcs2$, and $35\farcs4$ for 160, 250, 350, and 500 $\micron$ images, respectively. 
In particular, the 250 $\micron$ image is a good reference in comparison between the continuum and line emission because it has almost the same spatial resolution as that of our NRO 45-m observations.

\section{Results} \label{sec:res}

\begin{deluxetable*}{lCcCCDDDcCCDDD}
    \tablecaption{Line Statistics \label{tbl:result}}
    \tablehead{
    \colhead{} & \colhead{} & \multicolumn{8}{c}{\ctfs\ $(J=2-1)$} & \colhead{} & \multicolumn{8}{c}{\nthp\ $(J=1-0)$} \\
    \cline{4-11} \cline{13-20} \colhead{Core} & \colhead{$N_\mathrm{O}$} & \colhead{} & \colhead{$N_\mathrm{D}$} & \colhead{$\sigma_\mathrm{rms}$} & \multicolumn{2}{c}{$T_\mathrm{A}^*$} & \multicolumn{2}{c}{$V_\mathrm{LSR}$} & \multicolumn{2}{c}{$\Delta V$} & \colhead{} & \colhead{$N_\mathrm{D}$} & \colhead{$\sigma_\mathrm{rms}$} & \multicolumn{2}{c}{$T_\mathrm{A}^*$} & \multicolumn{2}{c}{$V_\mathrm{LSR}$} & \multicolumn{2}{c}{$\Delta V$} \\
    \colhead{} & \colhead{} & \colhead{} & \colhead{} & \colhead{[K]} & \multicolumn{2}{c}{[K]} & \multicolumn{2}{c}{[\kmps]} & \multicolumn{2}{c}{[\kmps]} & \colhead{} & \colhead{} & \colhead{[K]} & \multicolumn{2}{c}{[K]} & \multicolumn{2}{c}{[\kmps]} & \multicolumn{2}{c}{[\kmps]}}
    \decimals
    \startdata
    L1544  & 100 & & 62 & 0.03 & 0.13 &  7.09\pm0.01 & 0.37\pm0.01 & & 56 & 0.07 & 0.49\pm0.08 &  7.09\pm0.01 & 0.32\pm0.01 \\
    L1552  & 100 & & 22 & 0.05 & 0.10 &  7.63\pm0.02 & 0.43\pm0.05 & & 32 & 0.07 & 0.43\pm0.09 &  7.65\pm0.01 & 0.21\pm0.01 \\
    L1689B & 100 & & 72 & 0.04 & 0.24 &  3.53\pm0.01 & 0.34\pm0.01 & & 47 & 0.05 & 0.36\pm0.12 &  3.53\pm0.01 & 0.31\pm0.01 \\
    L694-2 & 100 & & 49 & 0.03 & 0.07 &  9.51\pm0.01 & 0.39\pm0.03 & & 49 & 0.05 & 0.44\pm0.09 &  9.54\pm0.01 & 0.28\pm0.01 \\
    L1197  &  81 & & 18 & 0.03 & 0.05 & -3.22\pm0.03 & 0.35\pm0.07 & & 23 & 0.04 & 0.25\pm0.12 & -3.19\pm0.01 & 0.27\pm0.01 \\
    \enddata
    \tablecomments{$N_\mathrm{O}$ and $N_\mathrm{D}$ is the number of the mapping points and the detected points, respectively. $\sigma_\mathrm{rms}$ is the noise level of the observations. $T_\mathrm{A}^*$, $V_\mathrm{LSR}$, and $\Delta V$ is the antenna temperature, the velocity position, and the FWHM line width of the averaged line profile for all detected points. $V_\mathrm{LSR}$ and $\delta V$ were measured by single or hyperfine Gaussian fittings for \ctfs\ and \nthp\ observations, respectively.}
\end{deluxetable*}

All the five targets were detected in both \ctfs\ $(2-1)$ and \nthp\ $(1-0)$ lines. 
The sensitivity level achieved in our observations in both \ctfs\ $(2-1)$ and \nthp\ $(1-0)$ lines is found to be $\sigma_\mathrm{rms}\approx 0.03-0.05$ K in a $\rm T^*_A$ scale. 
The \ctfs\ line shows a single Gaussian shape in all the cores except for a few spectra toward the central regions of L1544 and thus it is believed to be optically thin in most cases. 
As a result, it is highly unlikely that the distribution of CS molecules in all the targets is modified by any saturation effect in the line emission. 
The results of the \ctfs\ $(2-1)$ and \nthp\ $(1-0)$ line observations for all the targets are summarized in Table \ref{tbl:result}. 
The \ctfs\ and \nthp\ lines were detected at 46\% and 43\% of mapping points, respectively. 
The signal-to-noise (S/N) ratios in positions having strong detection are found to be $\sim 6$ for \ctfs\ and $\sim 10$ for \nthp. 
In the remainder of this work, we only considered the spectra with an S/N ratio greater than $2.5$.

\subsection{Line Profiles and Their Intensity Distributions} \label{sec:res-line}

\begin{figure}
    \centering
    \includegraphics[scale=0.85]{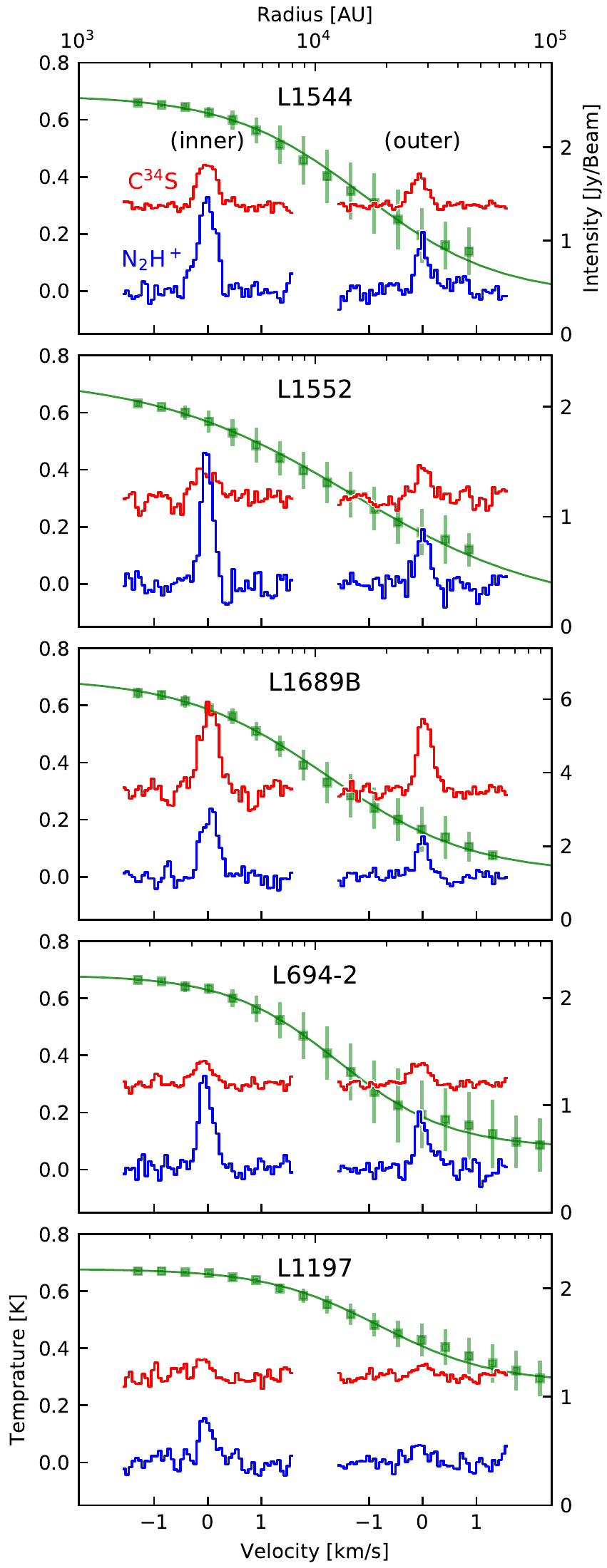}
    \caption{Radial variation of dust continuum and molecular line emission in five prestellar cores. The radial distribution of dust continuum drawn in green solid line is that of \textit{Herschel} 250 $\micron$ intensity. The radial distribution of the line emission intensity is expressed with two representative molecular line profiles in \nthp\ (in blue) and \ctfs\ (in red). The inner and outer line profiles are the average ones for the spectra within or outside an area of 70\% contour of the dust peak emission, respectively. \label{fig:line}}
\end{figure}

\begin{figure*}
    \centering
    \includegraphics[scale=0.9]{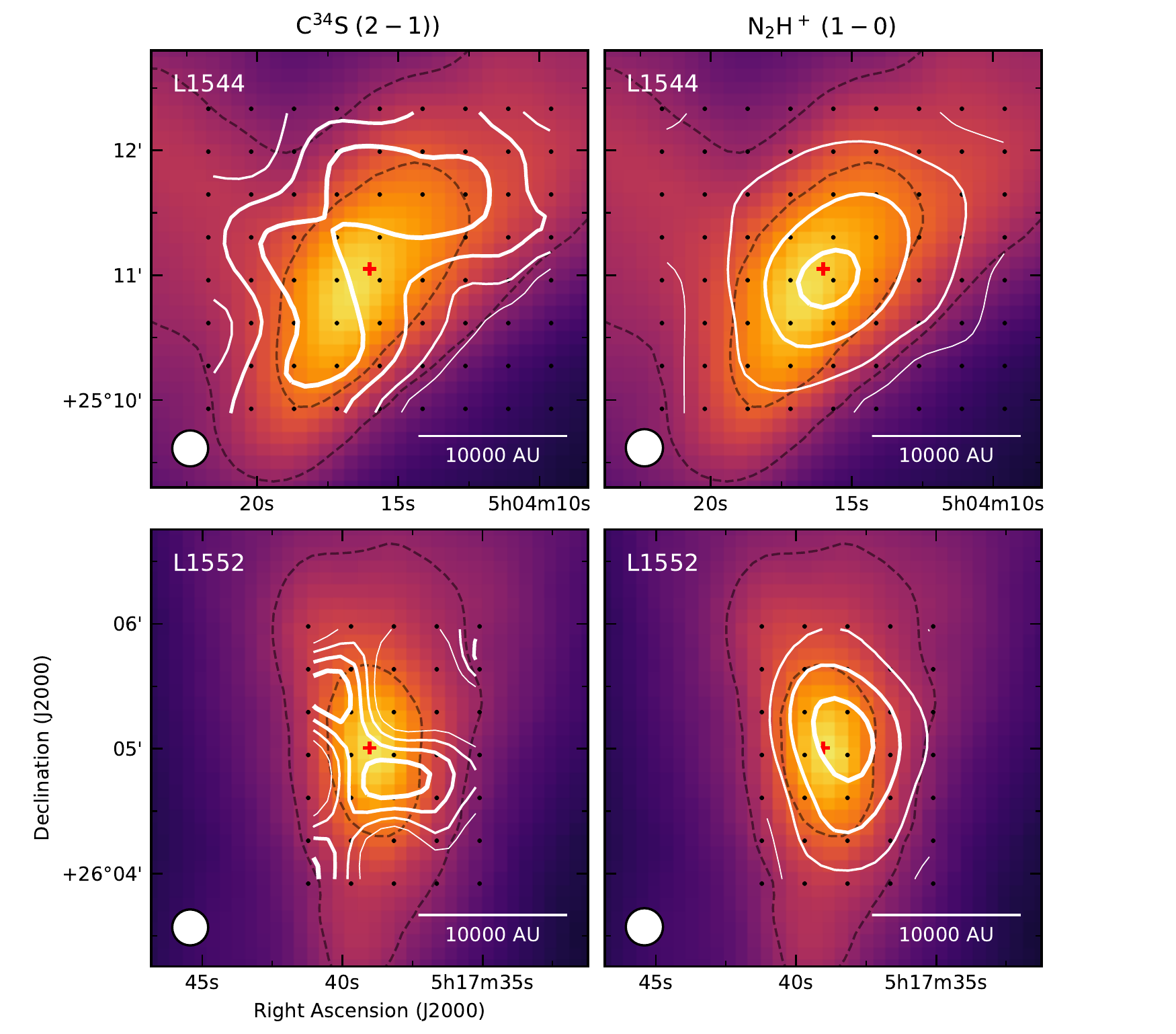}
    \caption{Comparison of the line intensity distribution of \ctfs\ (left panel) and \nthp\ (right panel) with 250 $\mathrm{\mu m}$ dust continuum distribution for our targets. 
    The continuum maps are given in color tone with two black dashed contours for 40\% and 70\% value of its peak intensity. Note that the 70\% contour was used as a boundary for inner and outer region in Figure \ref{fig:line}. The line intensity maps are drawn in several contours, the thinnest first contour indicating $3\sigma$ level of the integrated intensity, the thickest last contour for 90\% value of its peak integrated intensity, and the middle two contours evenly spaced between the first and last contour levels. Black dots in the maps are for the mapping points. Cross marks are the positions of the continuum intensity peaks whose coordinates are given in Table \ref{tbl:sample}. In each map, the size of the FWHM for NRO 45-m telescope and the scale bar of 10,000 au are displayed in left and right corners, respectively. \label{fig:map}}
\end{figure*}

\begin{figure*}
    \figurenum{2}
    \centering
    \includegraphics[scale=0.9]{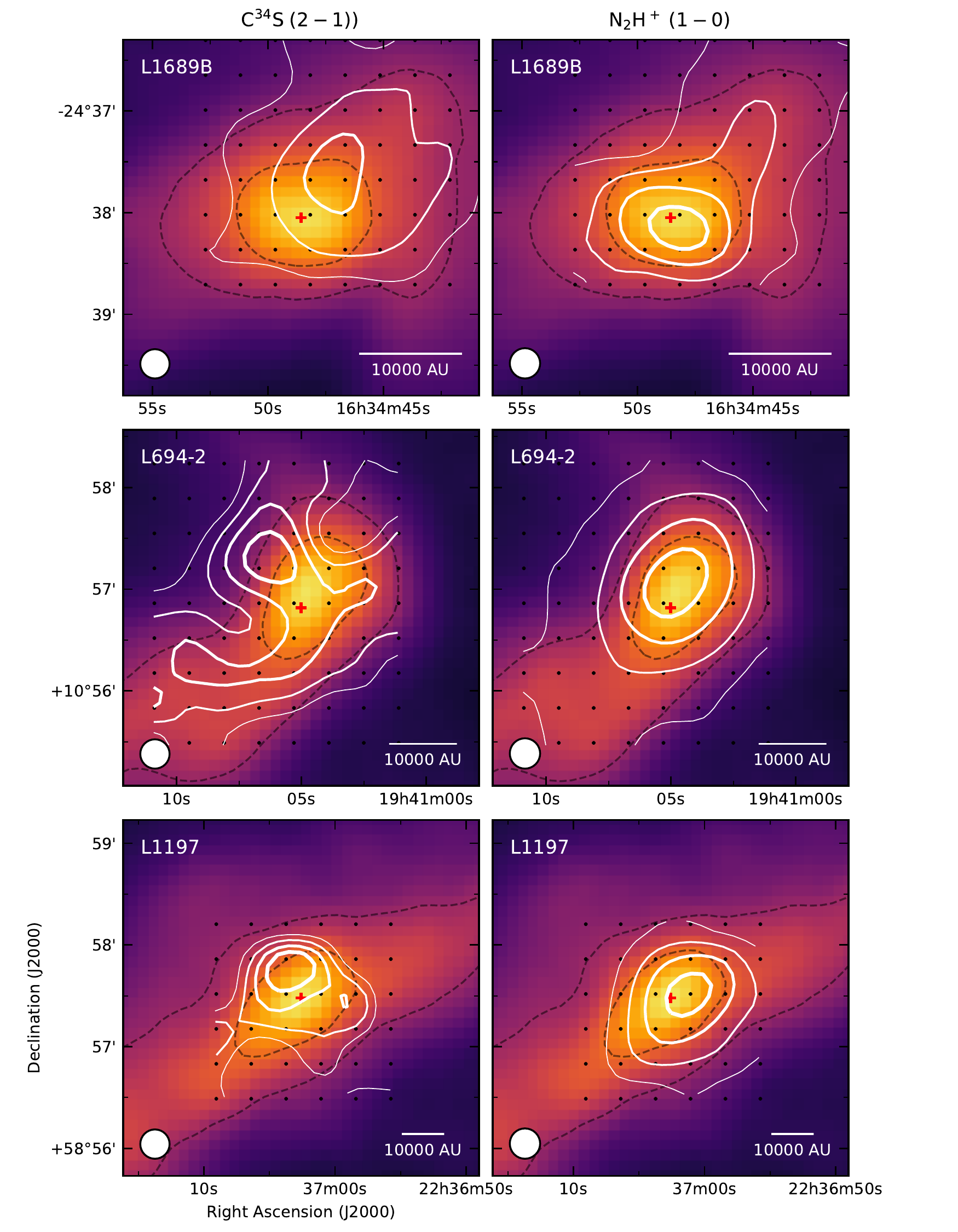}
    \caption{\textit{Continued}}
\end{figure*}
\setcounter{figure}{2}

The line profiles of \ctfs\ and \nthp\ are shown in Figure \ref{fig:line} with two averaged spectra for each line towards two different regions in each core, the ``inner'' region as the region encompassing 70\% of the dust peak emission and the ``outer'' region as the region outside the 70\% contour. 
Note that the area indicated by the 70\% contour on the dust intensity maps is almost identical to the central flat region in the starless cores. 
Radial distributions of \textit{Herschel} 250 $\micron$ dust continuum towards each target are also shown in Figure \ref{fig:line}. 
The most prominent feature seen in Figure \ref{fig:line} is that the dust continuum emission is relatively much brighter towards the inner region than the outside. 
Similar intensity variation is also seen in the case of \nthp\ which is known to not depleted even towards the highly evolved core \citep{Crapsi:2005kp,Bergin:2007iy}. 
It seems very natural that intensity of the optically thin and chemically unaffected \nthp\ line is brightened along with the density enhancement. 
However, the intensity variation seen between the inner and the outer regions in \ctfs\ is different in the sense that the line intensity is either slightly brighter in the inner region or remains fairly similar in both regions. 
As mentioned in Section \ref{sec:obs-line}, it is unlikely due to the saturation of the \ctfs\ line that the intensity of the \ctfs\ line does not increase as the intensities of continuum and \nthp\ line. 
The difference in the brightness distribution seen for \ctfs\ and \nthp\ lines as shown in Figure \ref{fig:line} suggests that \ctfs\ may be chemically affected in the central region of the dust continuum while \nthp\ may not be. 
This is in line with the expectations of previous studies on CS depletion \citep{Leger:1983wr, Bergin:1997fl, Tafalla:2002bn}.

The 1D radial variations in the molecular line intensities with respect to the dust continuum seen above can be further examined in 2D intensity maps. 
For this purpose, contours of the integrated intensities in \ctfs\ $(2-1)$ and \nthp\ $(1-0)$ lines are over-plotted on the \textit{Herschel} 250 $\micron$ dust continuum maps in color tones as shown in Figure \ref{fig:map}. 
It is quite apparent from the Figure \ref{fig:map} that both \nthp\ and the dust continuum intensities are getting stronger towards the central region in all the five cores studied here. 
In addition to that, the \nthp\ emission is singly peaked and coinciding with the dust continuum peak.

However, it is also clear that the distribution of \ctfs\ is very much different from that of the \nthp\ and the dust emission. 
The intensity peaks of \ctfs\ maps do not coincide with the peak positions of \nthp\ or dust emission maps. Furthermore, some of the \ctfs\ maps show multiple intensity peaks possibly due to the central depletion of CS molecule. 
These results further indicate that in all the five cores studies here, the CS molecules may be significantly depleted out in the central high density region while \nthp\ molecule may not be.

\subsection{Abundance Variation Along the Radii of the Cores} \label{sec:res-xmol}

\begin{figure*}
    \includegraphics[scale=0.9]{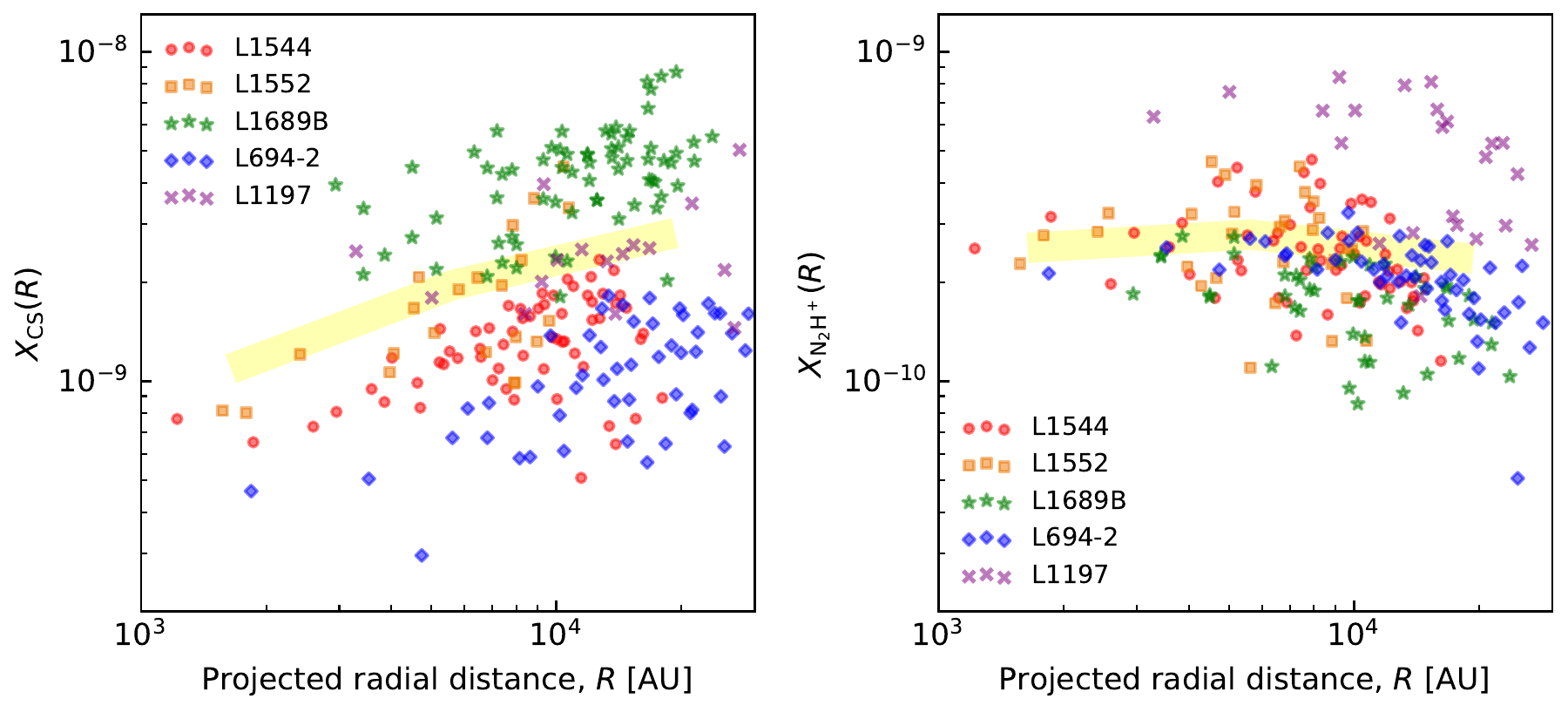}
    \centering
    \caption{Abundance variation of CS and \nthp\ molecules in prestellar cores as a function of their radial distance. Projected radial distance, $R$, is measured from the peak position of \hmol\ column density. The thick yellow lines indicate the average values for three intervals between the radial distances of $10^3$, $10^{3.5}$, $10^4$ and $10^{4.5}$ au. \label{fig:prof}}
\end{figure*}

The spatial variation of the molecular depletion can be examined more precisely if the ratio of its content with respect to the \hmol\ molecules (so called the molecular abundance) is estimated. 
Thus we made an attempt to obtain the abundances of CS and \nthp\ at each of the positions of our observations by dividing the column densities of each molecules by the column density of \hmol\ in the following manner;
\begin{equation}
	X(\mathrm{CS}) = \frac{N(\mathrm{CS})}{N(\mathrm{H_2})} = \frac{22.7\times N(\mathrm{C^{34}S})}{N(\mathrm{H_2})}\; ,
\end{equation}
and
\begin{equation}
	X(\mathrm{N_2H^+}) = \frac{N(\mathrm{N_2H^+})}{N(\mathrm{H_2})}\; ,
\end{equation}
where $X$ is the abundance, $N$ is the column density, and 22.7 is the solar isotopic abundance ratio between $\mathrm{C^{32}S}$ (main isotopologue of CS molecule) and $\mathrm{C^{34}S}$, $\mathrm{C^{32}S}/\mathrm{C^{34}S}$, which was found to be consistent with the ratio in the ISM \citep{Lucas:1998ue}.

The $N(\mathrm{H_2})$ and dust temperature ($T_\mathrm{d}$) were measured from the spectral energy distribution (SED) fit of the \textit{Herschel} dust emission at 160, 250, 350, and 500 \micron\ with the modified black-body function and the $N(\mathrm{N_2H^+})$, and $N(\mathrm{CS})$ were calculated from the integrated intensity of our \nthp and \ctfs\ molecular line observations. 
The entire procedure to derive the column densities is described in detail in Appendix \ref{app:cdcal}.

From the derived molecular and \hmol\ column densities at every pixel of our targets cores, we derived its corresponding abundance. 
We show the abundance values for all the sources in Figure \ref{fig:prof} as a function of the radial distances. 
This figure clearly indicates how the abundances of the two molecules change radially. 
First of all, it is apparent that the CS abundance decreases overall toward the central region of the dense cores. 
The average value of the CS abundance (yellow lines in Figure \ref{fig:prof}) of $\sim 2.8\times 10^{-9}$ obtained from the low-density outer regions ($> 10^4$ au) gets reduced to $\sim 9.6\times 10^{-10}$ in the high-density core center ($< \sim 3\times10^3$ au). 
Because the spatial resolution for the targets is greater than 1,000 au in the linear scale, the number of data points near the center is relatively low. 
Nonetheless, the decline of the CS abundance at the central regions of the cores is much more apparent in the radial abundance profile, about three times less than the abundance at the outer regions of the cores. 
Note that, in the case of L1197, it is difficult to ascertain a significant reduction in the abundance towards its central region. 
This is because L1197 is the most distant among our targets and thus there is no data point within 3,000 au towards the central region of the core to be used for any discussion on the CS depletion.

On the contrary, it is also clear that \nthp\ abundance is $\sim 2.3\times 10^{-10}$, almost constant over the whole radial distances. 
This suggests that \nthp\ molecule barely suffers from being frozen out at the central regions of the dense cores. 
The result is consistent with the previous studies which asserted that the N-bearing molecules survive at higher density regions in starless cores than the C-bearing species \citep{Bergin:1997fl,Caselli:2002cb}.

\section{Discussion} \label{sec:disc}

\subsection{Effects by Outer Envelope Lying in the Line-of-Sight in the Calculation of Molecular Abundance Profile} \label{sec:disc-under}

\begin{figure*}
    \centering
    \includegraphics[scale=1.0]{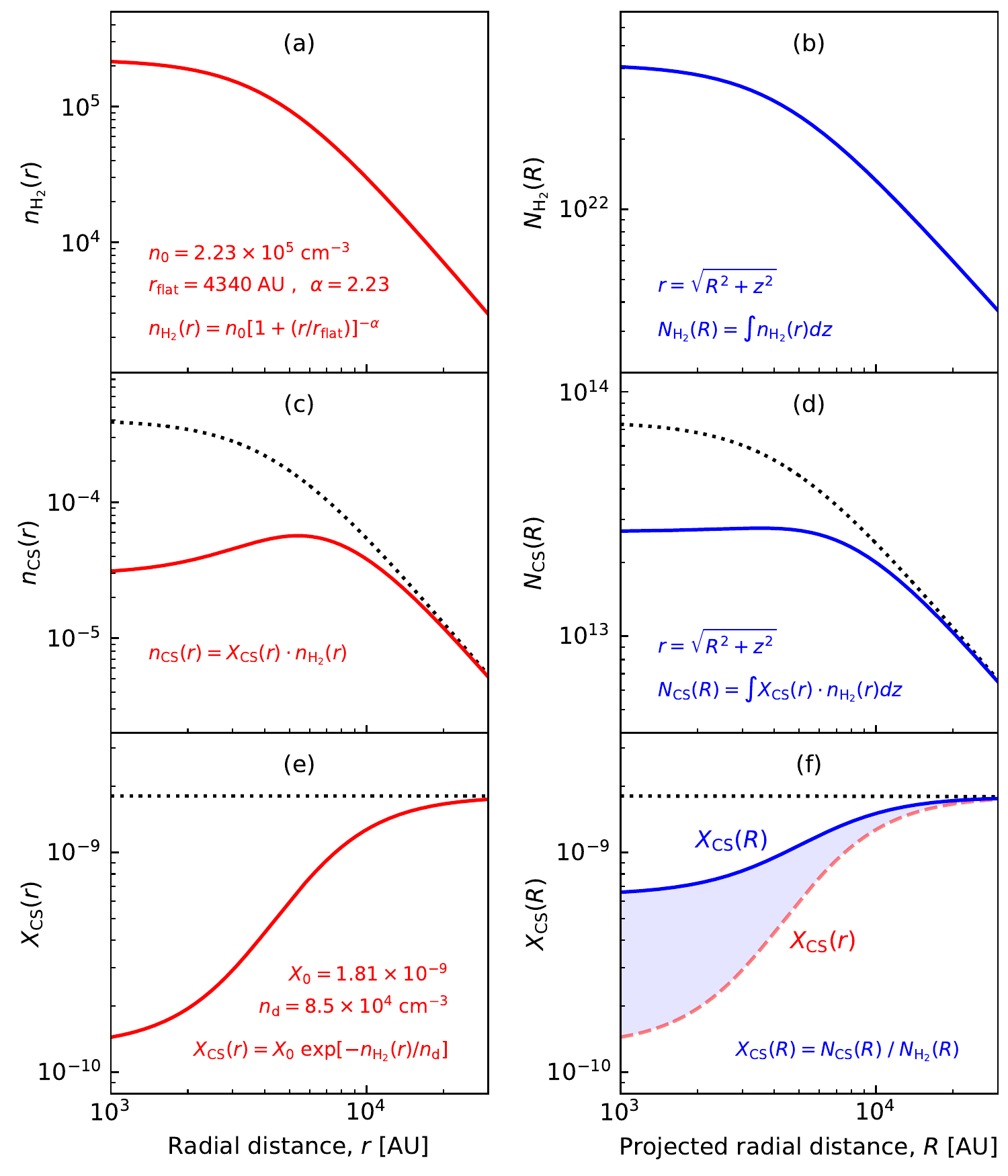}
    \caption{CS depletion expressed with two physical parameters, the volume density (left panels) and the column density (right panels). The $r$ and $R$ in the x axis designate the 3D radial distance from the center of a model prestellar core and the projected distance from the same center, respectively. The $n_\mathrm{H_2}(r)$, $n_\mathrm{CS}(r)$, and $X_\mathrm{CS}(r)$ in y axes of the left panels are the \hmol\ volume density, CS volume density and the corresponding abundance at $r$, respectively, while $N_\mathrm{H_2}(R)$, $N_\mathrm{CS}(R)$, and $X_\mathrm{CS}(r)$ in y axes of right panels are the \hmol\ column density, CS column density and the corresponding abundance at $R$, respectively. The volume density and abundance profiles shown at the left panels are given with a 3D density structure model for \hmol\ and CS using the parameters obtained from the analysis of L1544 by \citet{Tafalla:2002bn}. Similarly, column density profiles as a function of $R$ at the right panels are obtained by integrating $n(r)$ along the line-of-sight. The blue shaded area shows the difference between the actual CS abundance profile by the ratio of CS and \hmol\ volume densities and the other abundance profile estimated from the ratio of CS and \hmol\ column densities.}
    \label{fig:vdvscd}
\end{figure*}

\begin{deluxetable*}{lDDDcDDDcDD}
    \tablecaption{The best fit results for \hmol\ volume density and molecular abundance models. \label{tbl:mcmc}}
    \tablehead{
    \colhead{} & \multicolumn{6}{c}{\hmol\ volume density\tablenotemark{a}} & \colhead{} & \multicolumn{6}{c}{\nthp\ abundance\tablenotemark{b}} & \colhead{} & \multicolumn{4}{c}{CS abundance\tablenotemark{c}} \\
    \cline{2-7} \cline{9-14} \cline{16-19} \colhead{Core} & \multicolumn{2}{c}{$n_0$} & \multicolumn{2}{c}{$r_\mathrm{flat}$} & \multicolumn{2}{c}{$\alpha$} & \colhead{} & \multicolumn{2}{c}{$X_\mathrm{const.}$} & \multicolumn{2}{c}{$X_0$} & \multicolumn{2}{c}{$\beta$} & \colhead{} & \multicolumn{2}{c}{$X_\mathrm{out}$} & \multicolumn{2}{c}{$n_\mathrm{d}$} \\
    \colhead{} & \multicolumn{2}{c}{[$10^5\;\mathrm{cm^{-3}}$]} & \multicolumn{2}{c}{[$10^3$ AU]} & \multicolumn{2}{c}{} & \colhead{} & \multicolumn{2}{c}{[$10^{-10}$]} & \multicolumn{2}{c}{[$10^{-10}$]} & \multicolumn{2}{c}{} & \colhead{} & \multicolumn{2}{c}{[$10^{-9}$]} & \multicolumn{2}{c}{[$10^4\;\mathrm{cm^{-3}}$]}}
    \decimals
    \startdata
    L1544  & 2.2\;_{-0.3}^{+0.3}    & 4.3\;_{-0.8}^{+0.9}  & 2.2\;_{-0.2}^{+0.3} & & 2.68\;_{-0.03}^{+0.03} & 2.51\;_{-0.09}^{+0.09} & -0.03\;_{-0.01}^{+0.01} & & 1.81\;_{-0.04}^{+0.04} & 8.5\;_{-0.5}^{+0.6} \\
    L1552  & 2.1\;_{-0.3}^{+0.4}    & 3.3\;_{-0.6}^{+0.7}  & 2.3\;_{-0.2}^{+0.2} & & 2.77\;_{-0.04}^{+0.04} & 2.5\;_{-0.1}^{+0.1}    & -0.04\;_{-0.02}^{+0.02} & & 2.3\;_{-0.2}^{+0.2}    & 7.\;_{-2.}^{+2.} \\
    L1689B & 1.3\;_{-0.1}^{+0.2}    & 5.8\;_{-0.9}^{+0.9}  & 2.4\;_{-0.2}^{+0.3} & & 2.19\;_{-0.03}^{+0.03} & 2.3\;_{-0.1}^{+0.1}    & 0.03\;_{-0.02}^{+0.02}  & & 6.9\;_{-0.1}^{+0.2}    & 3.9\;_{-0.2}^{+0.2} \\
    L694-2 & 0.9\;_{-0.1}^{+0.1}    & 11.\;_{-2.}^{+2.}    & 2.8\;_{-0.5}^{+0.7} & & 1.88\;_{-0.02}^{+0.02} & 2.55\;_{-0.06}^{+0.07} & 0.24\;_{-0.02}^{+0.02}  & & 1.24\;_{-0.06}^{+0.06} & 3.6\;_{-0.3}^{+0.4} \\
    L1197  & 0.12\;_{-0.01}^{+0.01} & 13.\;_{-2.}^{+2.}    & 2.2\;_{-0.2}^{+0.3} & & 5.2\;_{-0.1}^{+0.1}    & 11.4\;_{-0.7}^{+0.8}   & 0.58\;_{-0.06}^{+0.06}  & & 2.12\;_{-0.09}^{+0.09} & - \\
    \enddata
    \tablecomments{The error of all the values represents $\pm 1\sigma$ range of the final posterior probability distribution from the MCMC analysis.}
    \tablenotetext{a}{$n(r) = n_0 [1+r/r_\mathrm{flat}]^{-\alpha}$, $n_0$ is the \hmol\ volume density at the core center, $r_\mathrm{flat}$ is the radius of the flat region, and $\alpha$ is the asymptotic power index.}
    \tablenotetext{b}{$X_\mathrm{const.}$ is the \nthp\ abundance in the constant abundance model, $X(r) = X_\mathrm{const.}$. $X_0$ is the \nthp\ abundance at the core center and $\beta$ is the power-law index in the centrally enhanced abundance model, $X(r) = X_0 [n(r)/n_0]^\beta$.}
    \tablenotetext{c}{$X_\mathrm{out}$ is the CS abundance without depletion at large radii and $n_\mathrm{d}$ is critical density of depletion in the CS depletion model, $X(r) = X_\mathrm{out} \mathrm{\;exp}[-n(r)/n_\mathrm{d}]$.}
\end{deluxetable*}

\begin{figure*}
    \centering
    \includegraphics[scale=0.85]{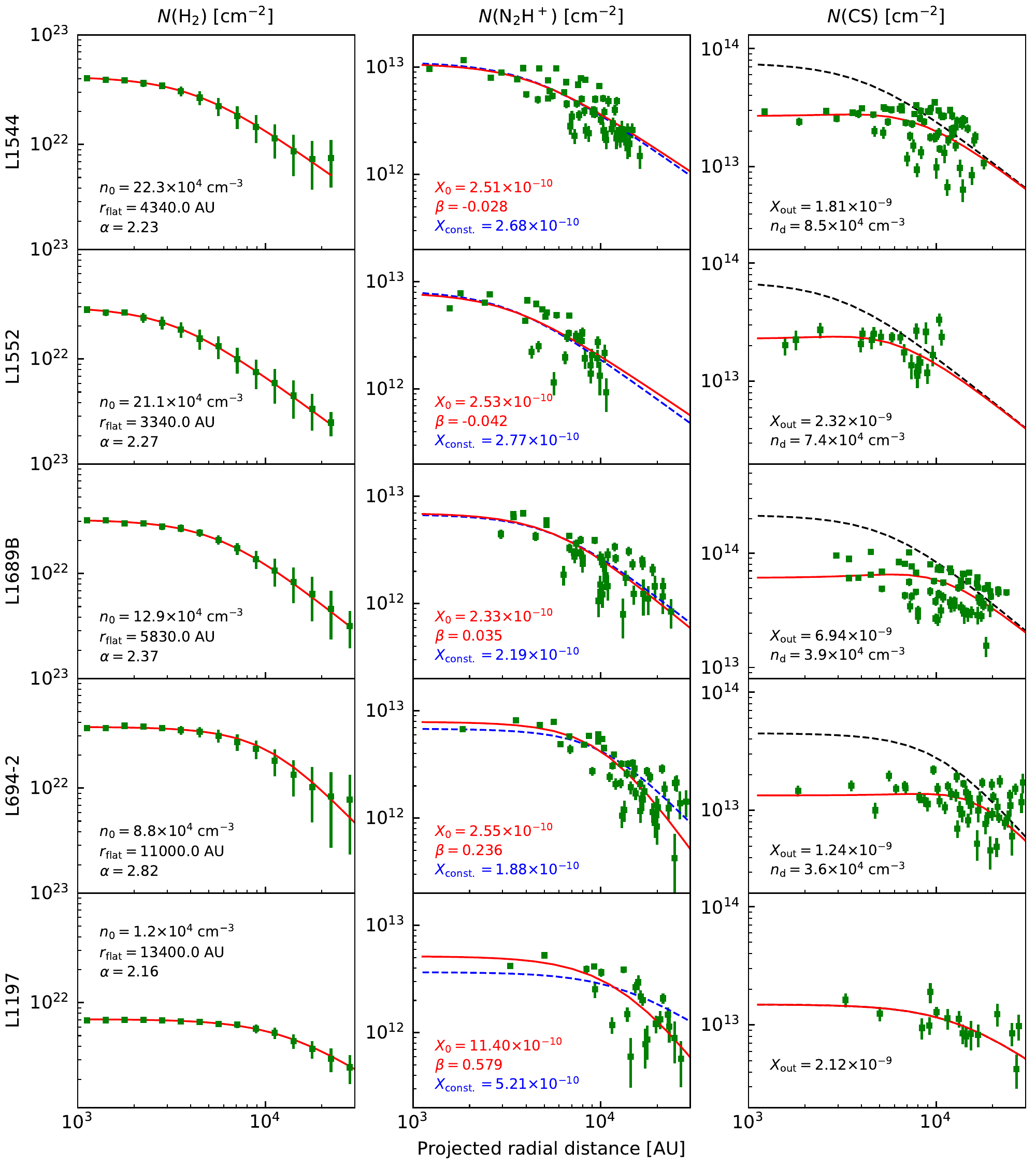}
    \caption{The radial column density profiles of \hmol, \nthp, and CS for our targets. The plots on the left panel are \hmol\ column density data from Herschel dust continuum emission (data points and their error bars in green) and \hmol\ column density profile derived by the centrally enhanced density model (red line). The best fit parameters for each target are given in each box panel. The plots in the middle panel are $N(\mathrm{N_2H^+})$ data and the derived $N(\mathrm{N_2H^+})$ profiles based on two models. The one model adopts the constant abundance (blue dashed line) while the other model deals with a power-law abundance given with $X(r) = X_0 [n(r) / n_0]^\beta$ (red solid line). The best fit parameters are also given for each target. The plots on the right panel show the data points of $N(\mathrm{CS})$ and the derived $N(\mathrm{CS})$ profiles based on two models. One model uses the constant abundance model (black dashed line) while the other model chooses the exponential abundance variation given with	$X(r) = X_\mathrm{out} \mathrm{\;exp}[-n(r)/n_\mathrm{d}]$ (red solid line). The best fit parameters are given for each target in the panel.}
    \label{fig:cdpro}
\end{figure*}

\begin{figure}
    \centering
    \includegraphics[scale=0.75]{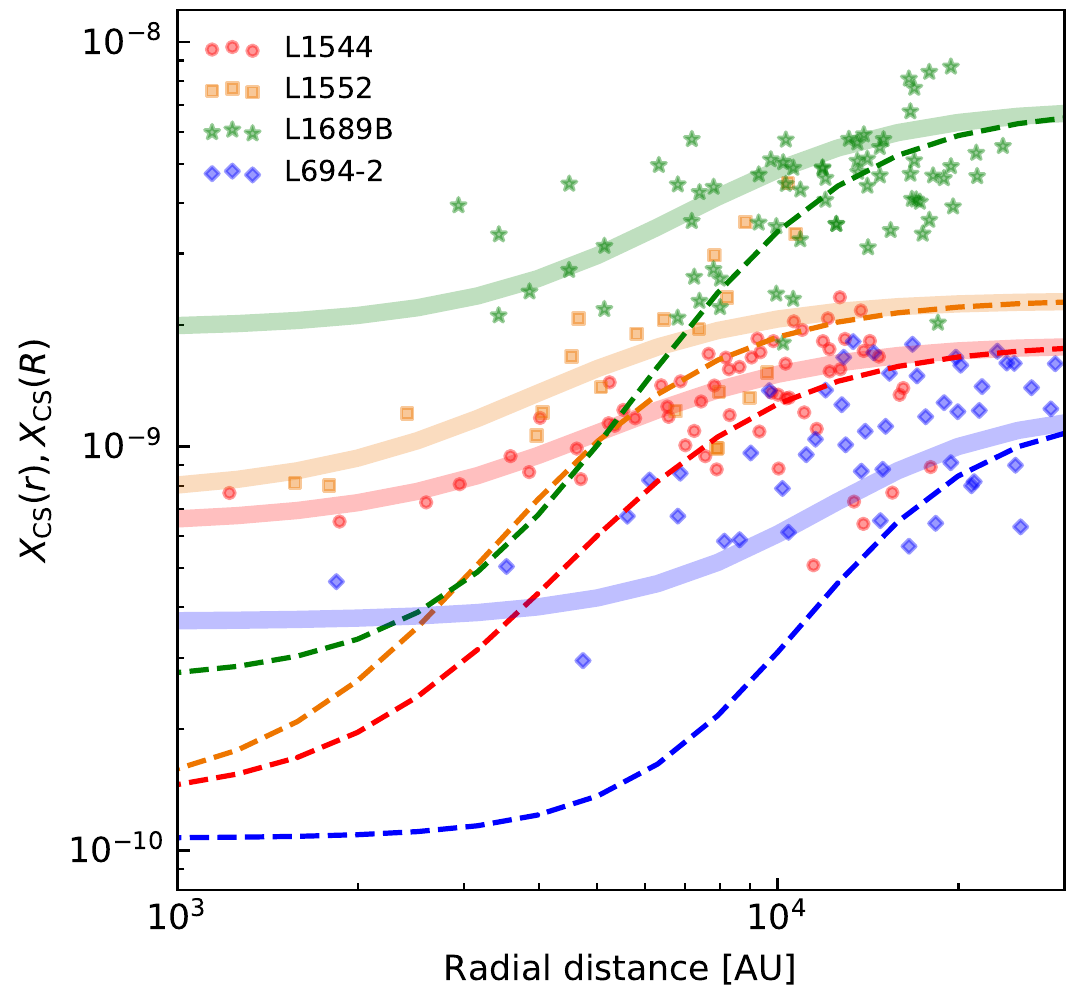}
    \caption{The radial abundance profiles by the best-fitted CS depletion models and the data for each target. The thick lines are the abundance profiles that are derived by column densities of CS and \hmol\ while the dashed lines are the other abundance profiles where the contamination effects by line-of-sight envelope components were minimized. The abundance profile for L1197, where the CS column density profile is well fitted by a constant abundance model without any depletion, is not drown in the plot. \label{fig:vdpro}}
\end{figure}

\begin{deluxetable*}{lDDDcDDD}
    \tablecaption{The degree of CS depletion before and after extracting the effect of the surrounding envelope. \label{tbl:depfac}}
    \tablehead{
    \colhead{} & \multicolumn{6}{c}{Based on column density\tablenotemark{a}} & \colhead{} & \multicolumn{6}{c}{Based on volume density\tablenotemark{b}} \\
    \cline{2-7} \cline{9-14} \colhead{Core} & \multicolumn{2}{c}{$X_\mathrm{in}$} & \multicolumn{2}{c}{$X_\mathrm{out}$} & \multicolumn{2}{c}{$f_\mathrm{D}$} & \colhead{} & \multicolumn{2}{c}{$X_\mathrm{in}$} & \multicolumn{2}{c}{$X_\mathrm{out}$} & \multicolumn{2}{c}{$f_\mathrm{D}$}}
    \decimals
    \startdata
    L1544      & $7.1\times10^{-10}$ & $1.7\times10^{-9}$ & 0.41 & & $1.8\times10^{-10}$ & $1.6\times10^{-9}$ & 0.11 \\
    L1552      & $8.9\times10^{-10}$ & $2.2\times10^{-9}$ & 0.40 & & $2.3\times10^{-10}$ & $2.2\times10^{-9}$ & 0.11 \\
    L1689B     & $2.1\times10^{-9}$  & $6.2\times10^{-9}$ & 0.33 & & $3.2\times10^{-10}$ & $5.7\times10^{-9}$ & 0.06 \\
    L694-2     & $3.7\times10^{-10}$ & $1.0\times10^{-9}$ & 0.39 & & $1.1\times10^{-10}$ & $8.1\times10^{-9}$ & 0.13 \\
    \hline
    On average & $1.0\times10^{-9}$  & $2.8\times10^{-9}$ & 0.36 & & $2.1\times10^{-10}$ & $2.6\times10^{-9}$ & 0.08 \\
    \enddata
    \tablecomments{$X_\mathrm{in}$ is an averaged CS abundance values of the inner region ($r<10^{3.5}$ AU) and $X_\mathrm{out}$ is an averaged CS abundance values of the outer region ($r>10^4$ AU). $f_\mathrm{D}$ is the abundance ratio of $X_\mathrm{in}/X_\mathrm{out}$ indicating the degree of CS depletion.}
    \tablenotetext{a}{CS abundance ratio was directly measured by the ratio of the CS and \hmol\ column density.}
    \tablenotetext{b}{To extract the effect of the presence of the CS molecules in the surrounding envelope along the line-of-sight, CS abundance ratio was derived from the CS and \hmol\ volume density which were inferred by using the core density structure and the CS abundance depletion models.}
\end{deluxetable*}

The degree of the CS depletion in the central region of a prestellar core can be examined by comparing the ratio of CS abundance at the central region to that at the outer region of the core using the expression 
\begin{equation}
    f_\mathrm{D} = \frac{\overline{X}_\mathrm{CS}(r<10^{3.5}\;\mathrm{au})}{\overline{X}_\mathrm{CS}(r>10^4\;\mathrm{au})} \; ,
\end{equation}
where $f_\mathrm{D}$ represents the degree of depletion. 
While the typical radius of the flat region of the starless core center is considered as the boundary of the central region, the typical size of the starless cores itself \citep{Bergin:2007iy} is taken as the boundary of the outer region. 
The boundary of the central and the outer regions in the case of L1544 are found to be $\sim 20\arcsec$ and $\sim 70\arcsec$ (expressed in linear scales), respectively. 
This is because CS will be depleted significantly at the central region while no such depletion may happen at the outer region. 
The ratio obtained in a similar way from the depletion model for five starless cores which include L1544 is found to be very small ($X(r=20\arcsec)/X(r=100\arcsec)\sim 0.01$) in a previous study conducted by \citet{Tafalla:2002bn}, implying that the CS depletion is highly severe in the central region of a prestellar core. 
However, in contrast, from our analysis on the five prestellar cores (Table \ref{tbl:depfac}), on averaging, obtained a much larger value ($f_\mathrm{D} \sim 0.4$).

One of the most plausible reasons for this difference could be that in the previous study volume densities were used to derive their abundance ratio while in our estimation we considered the column densities for CS and \hmol. 
In the former study \hmol\ and CS number densities were inferred by using the core density structure and the CS abundance depletion models \citep{Tafalla:2002bn}. 
Though this method was more straightforward in deriving the depletion factor, the results could be model-dependent. 
On the other hand, the ratio obtained using the column densities in our study might have got affected by the inclusion of outer envelope components along the line-of-sight where only small or no CS depletion effect may exist.

To verify this effect, the abundance ratio obtained using the volume densities of the central and outer envelope regions is compared with the ratios obtained from the column densities of the central and outer regions in Figure \ref{fig:vdvscd}. 
In the panel (f) of the Figure \ref{fig:vdvscd}, it is quite apparent that the abundance ratio obtained from the column densities can be significantly affected by the envelope component and hence becomes much greater than the abundance ratio obtained from the volume densities. 
Below we explain how we assumed the volume density profiles for CS and \hmol\ to calculate their column densities and abundances.

We began with the same models which were used by \citet{Tafalla:2002bn} to evaluate the core density and CS depletion as a function of the radial distance ($r$). 
The model for \hmol\ volume density at $r$ is given by
\begin{equation}
	n(r) = \frac{n_0}{1+(r/r_\mathrm{flat})^\alpha}\; ,
\end{equation}
where $n_0$ is the volume density at the core center, $r_\mathrm{flat}$ is the radius of the flat region, and $\alpha$ is the asymptotic power index.
The depletion model for the CS molecule was chosen according to the equation;
\begin{equation}
	X(r) = X_\mathrm{out} \mathrm{\;exp}\left[-n(r)/n_\mathrm{d}\right]\; ,
\end{equation}
where $X(r)$ is the molecular abundance at $r$ distance from the core center, $n_\mathrm{d}$ is the density giving an e-folding abundance drop, and $X_\mathrm{out}$ is the molecular abundance of very low density region at large radius. 
In this model the CS abundance is consistent with no depletion at the large radii but decreases exponentially toward the central region.
 
The \hmol\ and CS column densities are a total sum of the volume densities in unit column along the line-of-sight and thus can be expressed as: 
\begin{equation}
	N_\mathrm{H_2}(R) = \int n\left(\sqrt{R^2 + z^2}\right) dz\; ,
\end{equation}
\begin{equation}
	N_\mathrm{CS}(R) = \int X\left(\sqrt{R^2 + z^2}\right) \cdot n\left(\sqrt{R^2 + z^2}\right) dz\; ,
\end{equation}
where $R$ is the projected radial distance and $z$ is the depth of the line-of-sight from core center. 
These modelled column densities and the column densities estimated from the observational data for \hmol\ and CS were fitted with {\tt\string emcee}\footnote{\url{http://dfm.io/emcee/current/}} code which is the Markov Chain Monte Carlo (MCMC) sampler \citep{ForemanMackey:2013io}. 
This code allows the sampling to converge very efficiently to the target probability distribution when performing Bayesian analysis on high-dimensional optimization problems. 
By using this code, we were able to estimate the exact posterior probability distributions of the five parameters ($n_0$, $r_\mathrm{flat}$, $\alpha$, $X_\mathrm{out}$ and $n_\mathrm{d}$) rather than the simple least squares fitting results.

The results obtained from the fits are listed in Table \ref{tbl:mcmc} and shown in Figure \ref{fig:cdpro}. 
We found that both L1544 and L1552 have \hmol\ volume density of $\sim 2\times 10^5\;\mathrm{cm^{-3}}$, both L1689B and L694-2 have $\sim 1\times 10^5\;\mathrm{cm^{-3}}$, and L1197 has about $1.2\times 10^4\;\mathrm{cm^{-3}}$ at the core center. 
All the cores with a column density $>10^{22}$ cm$^{-2}$ showed a volume density of $>10^5$ cm$^{-3}$ at the core center. 
The size of the flat region, $r_\mathrm{flat}$, is found to be $\sim 4.9 \times 10^4$ au which corresponds to an FWHM of $\sim 0.05$ pc. 
The asymptotic power index is found to be $\sim 2.3$ on average.

We found that the CS abundance at the large radii showing no depletion is on average at $\sim 2.9\times 10^{-9}$, and the abundance $e$-folding density is $\sim 5\times 10^4\;\mathrm{cm^{-3}}$. 
The degree of CS depletion (as defined in Section \ref{sec:disc-under}) based on the volume density ratio of CS and \hmol\ is on average found to be $\sim 0.08$ for all the targets, 0.14 for a core with a central density of $10^5$ cm$^{-3}$, and 0.02 for a core with a central density of $2\times 10^5$ cm$^{-3}$.

Figure \ref{fig:vdpro} shows the radial variation of the CS abundances for each of the targets overlaid by the best fitted CS depletion models. 
The radial abundances given by the column densities containing the outer envelope indicate that the central depletion of each core appears rather mild, and probably affected by its envelope where little depletion is occurring. 
However, the radial abundances given by the CS volume densities clearly show that there must be a significant depletion of the CS molecule in the central region of each core. 
This depletion of the CS seems to be significantly occurring in all our targets except for the L1197.

In the case of L1197, the center density was found to be the lowest among our targets, as a result, any hint of CS depletion is barely seen. 
In fact the CS column density profile of L1197 is found to be well fitted by a constant abundance model without any depletion as seen in Fig. \ref{fig:cdpro}. 
Therefore it is likely that the central region of L1197 may not suffer from CS depletion as seriously as in the case of other targets. However, we note that the position where the \ctfs\ emission peaks does not coincide well with that of the dust continuum or \nthp\ emission. 
Additionally, relatively large distance of this object (farthest among our targets) makes it harder for us to see the details of the central region. 
Thus, in order to confirm whether L1197 is suffering from central depletion in CS or not at an equivalent level of spatial resolution to those of other targets, further observations in higher angular resolution are required. 

For the reference molecule, \nthp, we estimated the volume density profiles in the same manner as we carried out for the CS molecule. 
We applied two different forms of abundance profile model for the \nthp\ molecule. 
The first one is a constant abundance model in which we presume that the \nthp\ has not been significantly depleted over the entire density range of the starless cores as expressed by $X(r) = X_\mathrm{const.}$. 
The second one is a centrally enhanced model with respect to the outer low-density region of a dense core which is expressed by the equation 
\begin{equation}
    X(r) = X_0 \left[n(r)/n_0\right]^\beta\; ,
\end{equation}
where $\beta$ is a power-law index for $n(r)/n_0$. 
The subsequent analysis followed for the \nthp\ is same as that used for the CS. 
The fitting results for the two abundance models are also listed in Table \ref{tbl:mcmc} and shown in Figure \ref{fig:cdpro}. 
For the cores with the central volume density $>10^5$ cm$^{-3}$ (L1544, L1552 and L1689B), the results obtained from both the models are found to be similarly well fitted with the data points and the values of the $\beta$ obtained for the second (centrally enhanced) model are found to be close to zero. 
On the other hand, in the case of L694-2 and L1197, which have relatively low central densities, the constant abundance model does not precisely reproduce the sharp decrease in the column density of \nthp\ towards the outer regions (large radial distances) of the cores, as suggested by the results from the observations. 
For two cores, the model with a slightly enhanced abundance of \nthp\ at the center with respect to that at the outer regions of the cores appears to describe the observations better. 
One possible explanation for the existence of these two groups can be made by a role of CO molecule as a destroyer of \nthp\ \citep{Bergin:1997fl,Aikawa:2001bw}. 
In other words three cores are better fitted with a constant abundance model possibly because these three cores have enough CO depletion over the cores so that the \nthp\ may have been barely destroyed to survive during their evolution periods. 
On the other hand two other cores are not so evolved that the severe CO depletion may not have significantly occurred especially in the outer parts of the cores while CO depletion in the central parts of the cores was on going. 
In this environment \nthp\ molecule in the outer parts of the cores can be more destroyed with respect to the \nthp\ at the central regions of the cores. 
We postulate that these two cores may be in this chemical status. 
Another possible explanation is related to the LTE assumption used to estimate molecular column density, which is discussed in the following section.

\subsection{Effect by LTE Assumption in Calculation of Molecular Abundances} \label{sec:disc-lte}

In this study, it was assumed that our target cores are in the local thermodynamic equilibrium (LTE) and the excitation temperature of both CS and \nthp\ molecules in the calculations of molecular column densities are the same as the dust temperature from the \textit{Herschel} data (Equation \ref{eq:nmol}).
The dust temperatures of our targets were found to be decreasing toward the inner region of the cores, approximately $\sim 12$ K in the less dense outer region and $\sim 9$ K towards the denser center.
However, the \ctfs\ $(2-1)$ and \nthp\ $(1-0)$ critical densities at $\sim 10$ K are about $4.6\times 10^5$ and $2.4\times 10^5$ cm$^{-3}$, respectively \citep{Lique:2006,Tsitali:2015co}. 
Only the density in central region of our target cores was close to the critical density of these molecular lines, and most of the outer regions are significantly less dense than the critical density and thus probably far from the LTE condition for  both lines.

In the non-LTE condition, the excitation temperature of both molecules is expected to be considerably lower in most of the outer regions than the dust temperature we used in the calculation of molecular column density.
Therefore the LTE assumption could instead lead to an underestimation of molecular column density in the outer regions in comparison with the central regions of  the cores.

If we were able to use the realistic structure of the excitation temperature in the core, CS abundance in the outer region would have gotten relatively larger and the degree of CS depletion would have been more enhanced. 
Therefore it is clear that the degree of CS depletion in core center can be stronger than the levels we are showing in this study if the excitation temperature of CS line over the core is precisely informed and its value is adopted in the calculation of the column density.

For the same reason, the LTE calculation of the \nthp\ column density by adoption of dust temperature instead of the excitation temperature might result in its underestimation in most of the outer regions.

Indeed, as we mentioned in previous Section \ref{sec:disc-under}, the \nthp\ emission in L694-2 and L1197 was well fitted assuming a centrally enhanced abundance. 
This is probably because two dense cores, L694-2 and L1197 have such outer regions where LTE condition is not very well satisfied for \nthp\ line.
On the other hand, it is noted that the \nthp\ emission in L1544, L1552, and L1689B was well explained without a central enhancement of the \nthp\ abundance. 
This may be because these cores are more dense in overall than the two cores and thus \nthp\ $(1-0)$ line is closely thermalized to apply the LTE approximation for its calculation of column density.

\subsection{Relative Evolutionary Ranks of Cores by Their Physical Properties} \label{sec:disc-evol}

\begin{deluxetable*}{lclclclclcl}
    \tablecaption{Relative Ranking for Examining Evolutionary Status of Cores\label{tbl:rank}}
    \tablehead{
    \multicolumn{1}{c}{Properties} & \multicolumn{2}{l}{L1544} &  \multicolumn{2}{l}{L1552} & \multicolumn{2}{l}{L1689B} & \multicolumn{2}{l}{L694-2} & \multicolumn{2}{l}{L1197}}
    \startdata
    $N(\mathrm{H_2})$ [$\times 10^{22}$ cm$^{-2}$]     & 1 & (4.2)  & 4 & (3.0)  & 3 & (3.1)  & 2 & (3.9)  & 5 & (0.7) \\
    $N(\mathrm{N_2 H^+})$ [$\times 10^{12}$ cm$^{-2}$] & 1 & (12)   & 4 & (7.8)  & 3 & (8.0)  & 2 & (8.2)  & 5 & (5.3) \\
    $n(\mathrm{H_2})$ [$\times 10^4$ cm$^{-3}$]        & 1 & (22)   & 2 & (21)   & 3 & (13)   & 4 & (8.8)  & 5 & (1.2) \\
    $r_{70}$ [$\times 10^3$ AU]                        & 2 & (4.0)  & 1 & (3.0)  & 3 & (5.0)  & 4 & (8.6)  & 5 & (12.6) \\
    $\Delta V_{\mathrm{N_2 H^+}}$ [km s$^{-1}$]        & 2 & (0.33) & 4 & (0.25) & 1 & (0.42) & 3 & (0.29) & 5 & (0.24) \\
    \hline
    Sum of Ranks & \multicolumn{2}{l}{7} & \multicolumn{2}{l}{15} & \multicolumn{2}{l}{13} & \multicolumn{2}{l}{15} & \multicolumn{2}{l}{25} \\
    \enddata
    \tablecomments{These physical properties are intended to indicate evolutionary status of starless cores\citep{Crapsi:2005kp}. $r_{70}$ is the radius of 70\% contour of the dust peak and $\Delta V_{\mathrm{N_2 H^+}}$ is the FWHM line width of \nthp\ emission in peak position. The numbers in Column 2 to 6 imply the rank of evolved status of each core using values of the physical quantities in the first column. The lower the sum, the more evolved the core.}
\end{deluxetable*}

\begin{figure}
    \centering
    \includegraphics[scale=0.8]{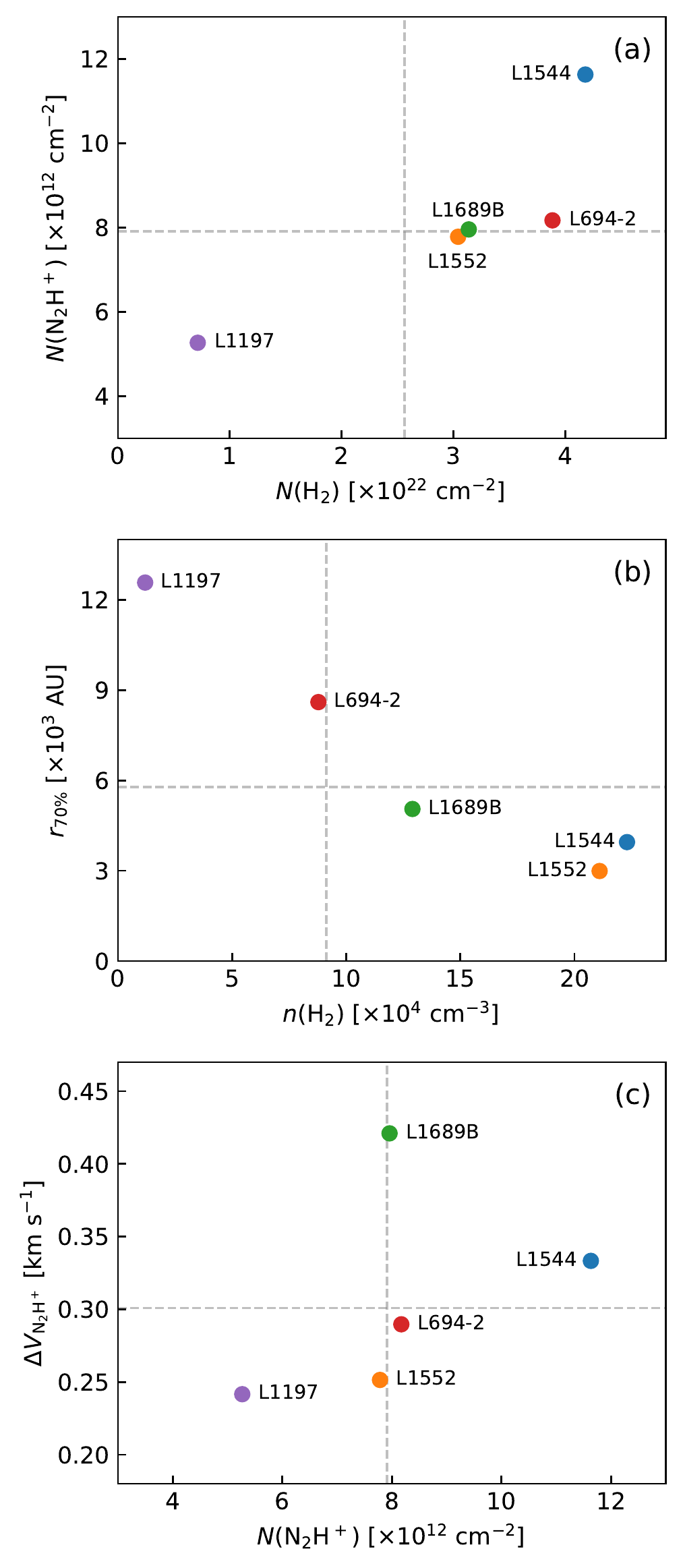}
    \caption{Relations among the physical quantities of our targets. (a) The \hmol\ column density and the \nthp\ column density at the peak position. (b) The \hmol\ volume density in core center and the radius of 70\% contour of the dust peak. (c) The \nthp\ column density and the FWHM line width of \nthp\ emission at the peak position. The dashed lines in (a), (b), and (c) are indicating the arithmetic mean values for our targets.
    \label{fig:evol}}
\end{figure}

\begin{figure}
    \centering
    \includegraphics[scale=0.8]{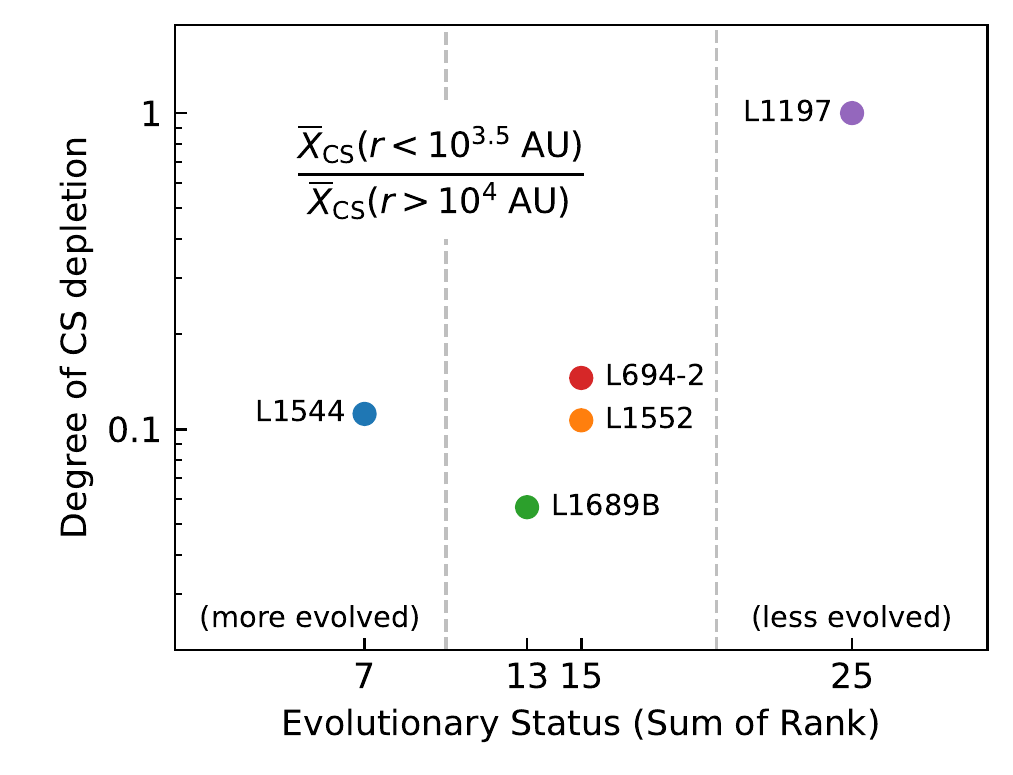}
    \caption{Comparison of the degree of CS depletion as a function of the ranking of relative evolution of cores (given in Table \ref{tbl:rank}). The degree of CS depletion, $\overline{X}(r<10^{3.5}\;\mathrm{au})/\overline{X}(r>10^4\;\mathrm{au})$, is determined from the abundance ratio of inner to outer region obtained by the inferred volume densities. The dashed line separates the three parts into a relatively more evolved, a less evolved, and an intermediate groups.
    \label{fig:evol-csdep}}
\end{figure}

In previous discussions we particularly focused on the regional difference of the CS depletion within the cores, finding more depletion at the central high density region than at the outer low density region. 
In this section we discuss on how the degree of CS depletion would vary from one core to another especially depending on its evolutionary status. 
One of the most important factors in our target selection regarding the core evolution was the presence of an extended infall asymmetry in the line profile. 
This selection criterion would ensure that the cores that we selected are all substantially evolved. 
What else would be an indicator for the core evolution? \citet{Crapsi:2005kp} have suggested several physical quantities of the cores as their evolution indicators such $N(\mathrm{N_2D^+})$, $N(\mathrm{N_2H^+})$, $[\mathrm{N_2D^+}]/[\mathrm{N_2H^+}]$, CO depletion factor, $n(\mathrm{H_2})$, $\Delta V_\mathrm{N_2H^+}$, skewness of infall asymmetry, and the radius of 70\% contour of the dust peak ($r_{70}$). 
They set the threshold value for eight physical quantities as an indicator of the evolved status for their target cores. 
For each physical quantity, one point was awarded if the value was greater than the threshold, and zero otherwise. 
They interpreted that the larger the total score (sum of all the individual points), the more evolved the core.

In this study we attempted to classify the evolved status in five prestellar cores studied here using five parameters that can be measured from our data. 
The parameters we used are the \hmol\ and \nthp\ column densities at peak position ($N(\mathrm{H_2})$ and $N(\mathrm{N_2H^+})$), the \hmol\ volume density at the core center ($n(\mathrm{H_2})$), the radius of 70\% contour of the dust peak ($r_{70}$), and the FWHM line width of \nthp\ emission at peak position ($\Delta V_\mathrm{N_2H^+}$). 
The line width of \nthp\ was measured by hyperfine fit for the seven components using the CLASS program with the assumption that all the components have the same excitation temperature and line width. 
Radius corresponding to $70\%$ contour of the dust peak indicates equivalent radius of the flattened region and was defined as $(a_{70} /\pi)^{1/2}$ with $a_{70}$ as the area within the $70\%$ contour in the \hmol\ column density maps.

The prestellar cores selected for the present study are all supposed to be highly evolved. 
We use similar criteria to \citet{Crapsi:2005kp} to infer the relative evolutionary status of the different cores in our sample. 
Therefore the parameters of our target prestellar cores are relatively compared and ranked. 
According to our scheme, the smaller the net sum of the ranks, the more evolved the core, and vice versa. 
The results are tabulated in Table \ref{tbl:rank} and are shown in Figure \ref{fig:evol}. 
Among our targets, L1544 is the most evolved and L1197 is the least evolved. 
L1689B, L1552, and L694-2 have quite similar total rank values. 
In the case of L1689B, all of the properties are lower-ranked than the arithmetic means while the \nthp\ line width at the position of peak emission is the most highly ranked among our targets. 
Between L1552 and L694-2, it is hard to determine the one which is more evolved. 
L1552 has a very high \hmol\ volume density in the core center, and the $r_{70\%}$ value is also very small. 
This indicates that the mass distribution of L1552 is concentrated towards the center. 
However, the \nthp\ column density and line width of L1552 are relatively low. 
On the other hand, L694-2 show a low \hmol\ volume density at the core center and a high value of $r_{70\%}$, but the \hmol\ and \nthp\ column densities are the second highest value among all our targets. 
Cores with higher volume densities may be considered as more evolved than those with higher column densities. 
But it is difficult to judge their relative importance because the volume density can be somehow model-dependent, while the column density is measured directly from the observations. 

Therefore our targets are classified into three groups as shown in Figure \ref{fig:evol-csdep} where the degree of CS depletion with respect to the evolutionary status (sum of the ranks) of cores is shown. 

L1197 which belong to the least evolved group shows no clear hint of CS depletion while the other four targets belonging to the most or intermediately evolved groups indicate significant CS depletion. 
Additionally, we found no significant difference in the degree of CS depletion among the four cores in these two groups. 
Probably it may be possible that all the cores except L1197 have already reached to the most evolved status and hence are indistinguishable based on the degree of CS depletion in them.

In fact, based on the theoretical study conducted by \citet{Bergin:1997fl}, it was reported that the abundance of sulfur-bearing molecules like CS is highly sensitive to the increase in the density. 
The CS depletion is expected to be low during the early stages of the core evolution ($n_\mathrm{H_2}\sim 10^4$ cm$^{-3}$) but gets severe as the core evolves. 
The CS molecules are shown to get significantly depleted at the collapsing phase before the initiation of the star formation ($n_\mathrm{H_2} \gtrsim 10^5$ cm$^{-3}$). 
All the cores, except for L1197, have the central density $n_\mathrm{H_2}\gtrsim 10^5$ cm$^{-3}$, suggesting that the CS molecules in the core center have already been sufficiently depleted. 
This is consistent with our results, as shown in Figure \ref{fig:evol-csdep}.

\section{Conclusion} \label{sec:conc}

We performed mapping observations of five prestellar cores, L1544, L1552, L1689B, L694-2, and L1197 in \ctfs\ $(2-1)$ and \nthp\ $(1-0)$ lines. 
The main aim of the study was to investigate the distribution of the CS emission throughout prestellar cores, in particular to study the CS depletion in prestellar cores, with respect to the inner and outer regions of each individual core and also with respect to the different evolutionary status of the cores. 
For this purpose we constructed the radial abundance profile of the CS and \nthp\ for each of the cores. The abundance profiles of the two molecules were derived using two different methods. 
In the first method we compared the observed molecular column densities of the CS and \nthp\ with the \hmol\ column density (from the \textit{Herschel} data). 
In the second one, we compared the volume densities of the CS, \nthp, and \hmol\ inferred from the fit performed on the observed column densities using the centrally enhanced density model. 
The results obtained from our observations and the analysis are summarized below:

1. In all our targets, the \ctfs\ emission intensity of inner regions is either comparable or weaker than that found towards the outer regions of the cores. 
The \ctfs\ integrated intensity maps of all the five cores show that either the maps are not centrally peaked or that the position where the intensity peaked is significantly shifted when compared with the distribution of \nthp\ line or dust continuum maps both of which showed similar single peaked distribution in the central region of the cores. 
The distribution of the \ctfs\ line emission suggests that the CS molecules got depleted significantly in the central high-density region of prestellar cores.

2. Such significant reduction of the CS molecules towards the core center is also seen in the radial profile of the CS abundance. 
The average value of the depletion factor between the central region ($\gtrsim 3,000$ au) and outer region ($\lesssim 10,0000$ au) of all the five cores is about 0.34. 
In other words the CS abundance is $\sim 2.8\times 10^{-9}$ at the outer region, but gets reduced by a factor of $\sim 3$ to $\sim 9.6\times 10^{-10}$ towards the core center due to its depletion. 
In contrast, \nthp\ abundance remains almost constant or slightly enhanced towards the central high-density region.

3. The CS abundances measured from the observed column densities of \hmol\ and CS are found to be affected by the presence of the molecules in the surrounding envelope overlapped to the line-of-sight. 
The depletion factor after extracting this effect using the centrally enhanced density model is found to be more significant as 0.06 to 0.11.

4. In calculation of molecular column density we assumed LTE condition with dust temperature over the core which decreases toward the central region of the core.
However, it is noted that most of the outer regions are less dense than the critical density for \ctfs\ $(2-1)$ and thus our LTE assumption can result in its underestimation in the outer region of the core, in comparison with more realistic situation where excitation temperatures can be adopted in the calculation of the column density. 
Therefore it is possible that the real degree of CS depletion in the central region of the core center can be stronger than we measured.

5. From our analysis of molecular lines and continuum data all cores except for L1197 are found to significantly suffer from CS depletion at the central region of the cores. 
In the case of L1197, the \ctfs\ emission distribution is likely indicative of CS depletion. 
However, its observed column density profile can be also explained with a constant radial abundance distribution without CS depletion at the central position. 
This is partly because of its flatter central \hmol\ column density distribution than other cores so that the density of L1197 is not as high as other cores to significantly cause the CS depletion or because of its larger distance from us than other cores so that there are little data to look for any strong evidence for CS depletion at the central region in our observing resolution.

6. We were able to group the cores according to the relative evolutionary status of the cores using five observed physical parameters such as the \hmol\ and \nthp\ column densities at their intensity peak position, the \hmol\ volume density at the core center, the radius of $70\%$ contour of the dust peak emission, and the FWHM line width of \nthp\ emission at dust peak position. 
L1544 and L1197 are classified to be the most evolved and the least evolved cores in our sample, respectively. 
The other cores, namely, L1552, L1689B and L694-2 are found to be at intermediate stages of their evolution. 
Although our statistics for CS depletion in the cores as a function of their evolution is highly limited, its is clear that four cores in more evolved status show a very significant depletion factor (0.06 - 0.13) while one core (L1197) in least evolved stage shows the least or no CS depletion. 

\acknowledgments
We acknowledge an anonymous referee for her/his careful reading of our manuscript to give us useful comments with which our paper was well improved.
This work was supported by Basic Science Research Program through the National Research Foundation of Korea (NRF) funded by the Ministry of Education, Science and Technology (NRF-2019R1A2C1010851).

\vspace{5mm}
\facilities{No:45m (BEARS), Herschel (PACS, SPIRE)}
\software{NEWSTAR, CLASS, astropy, scipy, emcee}

\appendix
\section{Estimation of {\hmol} column density and dust temperature}\label{app:cdcal}

\begin{deluxetable*}{lDDDDcDDDD}
    \tablecaption{Fitting results of radial intensity profiles for all {\textit{Herschel}} images \label{tbl:dustfit}}
    \tablehead{
    \colhead{} & \multicolumn{8}{c}{$I_0$ [$\mathrm{MJy\;sr^{-1}}$]} & \colhead{} & \multicolumn{8}{c}{$I_\mathrm{B}$ [$\mathrm{MJy\;sr^{-1}}$]} \\
    \cline{2-9} \cline{11-18} \multicolumn{1}{l}{Band [$\micron$]} & \multicolumn{2}{c}{$160$} & \multicolumn{2}{c}{$250$} & \multicolumn{2}{c}{$350$} & \multicolumn{2}{c}{$500$} & \colhead{} & \multicolumn{2}{c}{$160$} & \multicolumn{2}{c}{$250$} & \multicolumn{2}{c}{$350$} & \multicolumn{2}{c}{$500$}}
    \decimals
    \startdata
    L1544  &  89.6 & 190.9 & 193.1 & 116.6 & &   2.7 &  40.3 & 23.7 & 10.0 \\
    L1552  &  59.3 & 153.8 & 164.8 &  96.9 & & -10.9 &  42.4 & 25.2 & 11.0 \\
    L1689B & 310.6 & 484.5 & 358.2 & 165.2 & & 118.2 & 103.0 & 55.3 & 21.9 \\
    L694-2 & -     & 171.6 & 178.7 & 100.9 & & -     &  32.3 & 15.3 &  5.8 \\
    L1197  &  83.5 & 124.1 &  91.4 &  40.1 & &  -5.6 &  79.8 & 50.5 & 22.8 \\
    \enddata
    \tablecomments{$I_0$ is the central peak intensity without background and $I_\mathrm{B}$ is the background intensity caused by surrounding envelope and large-scale cloud emission in function as $I(R)=I_0/[1+(R/R_\mathrm{flat})^\alpha]+I_\mathrm{B}$.}
\end{deluxetable*}

\begin{deluxetable}{ccc}
    \tablecaption{Parameters for molecular column density calculation\label{tbl:const}}
    \tablehead{
    \colhead{Parameters} & \colhead{\ctfs} & \colhead{\nthp} \\
    \colhead{} & \colhead{$(J=2-1)$} & \colhead{$(J=1-0)$}}
    \startdata
    $\lambda\;[\mathrm{mm}]$ & 3.109563 & 3.217477 \\
    $A_{ul}$ & $1.600\times 10^{-5}$ & $3.628\times 10^{-5}$ \\
    $g_l / g_u$ & $3/5$ & $1/3$ \\
    $T_{\mathrm{ex}}\;\mathrm{[K]}$ & \multicolumn{2}{c}{$T_\mathrm{d}$\tablenotemark{a}} \\
    $T_{\mathrm{bg}}\;\mathrm{[K]}$ & \multicolumn{2}{c}{2.73} \\
    $B\;\mathrm{[MHz]}$ & 24103.541 & 46586.867 \\
    $Q_{\mathrm{rot}}(\mathrm{10\;K})$\tablenotemark{b} & 9.0 & 4.8 \\
    $E_l\;\mathrm{[K]}$ & 2.31 & 0 \\
    \enddata
    \tablenotetext{a}{From the results of the SED fit for \textit{Herschel} dust continuum data}
    \tablenotetext{b}{These values are examples at 10 K. We use the value obtained from $T_\mathrm{d}$ for each pixel.}
\end{deluxetable}

\begin{figure*}
    \centering
    \includegraphics[scale=0.9, angle=90]{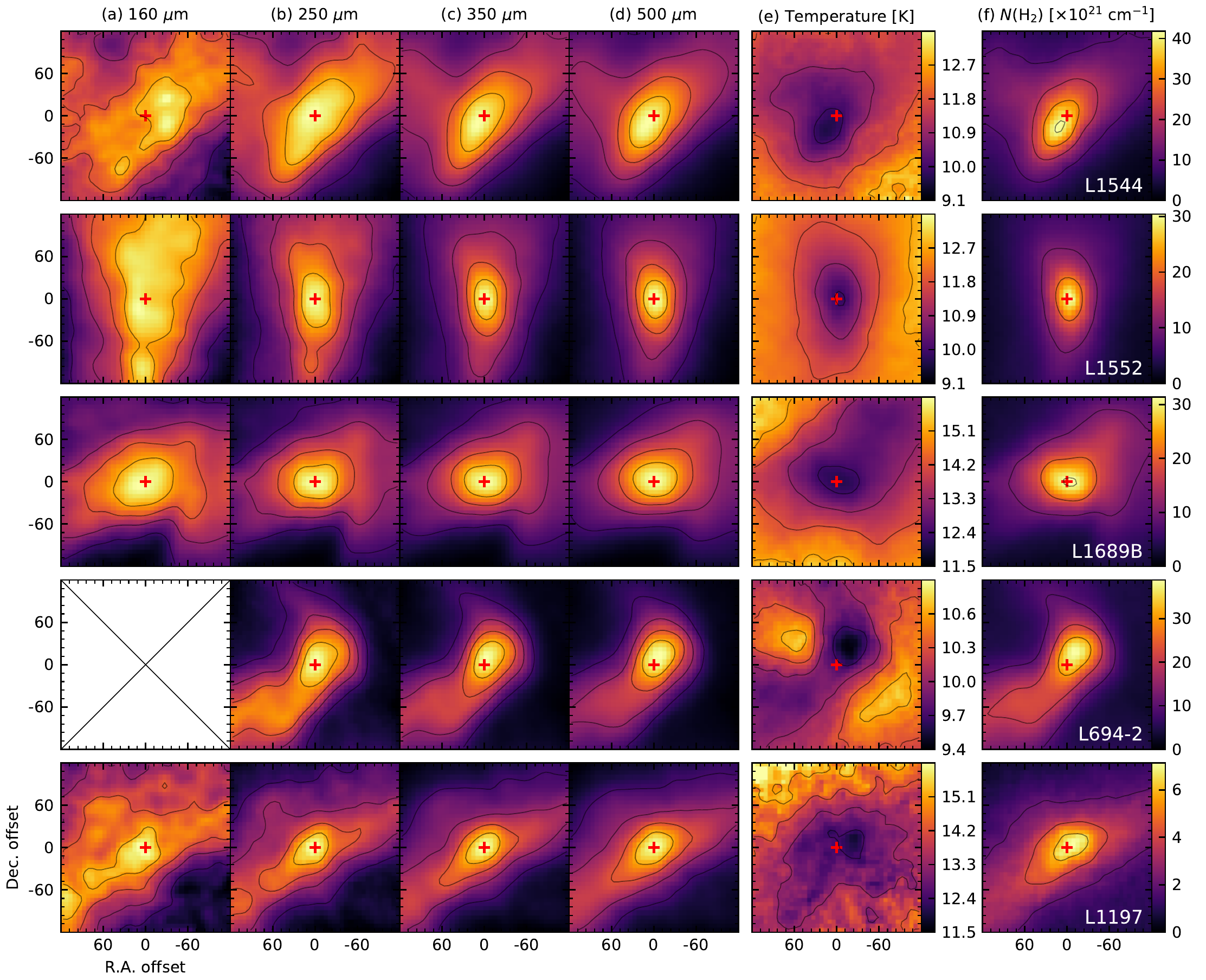}
    \caption{\textit{Herschel} images of five prestellar cores at 160 (a), 250 (b), 350 (c), and 500 \micron\ (d), and their dust temperature (e) and \hmol\ column density (f) maps. The dust temperature and \hmol\ column density maps were obtained from the SED fit of dust emission at 160 to 500 \micron. Black contours show 20, 40, 60, and 80\% levels of the maximum value in each \textit{Herschel} maps and show same levels with tick levels of scale bar in the dust temperature and \hmol\ column density maps. Red cross markers indicate a peak position of 250 \micron\ image.
    \label{fig:sedimage}}
\end{figure*}

\subsection{\hmol}\label{app:cdcal-hmol}

The $N(\mathrm{H_2})$, column density of \hmol\ gas is determined based on the spectral energy distribution (SED) fit of the \textit{Herschel} dust continuum data using the {\tt\string curve\_fit} function of the python package {\tt\string scipy}\footnote{\url{http://www.scipy.org/}}. 
The \textit{Herschel} data set consists of 160, 250, 350, and 500 $\micron$ images. 
The large \textit{Herschel} maps were trimmed to obtain images of appropriate sizes to include the region containing cores and exclude the surrounding extended cloud structures. 
The trimmed images were convolved with an FWHM of 17\farcs6, and re-gridded to have the same pixel size ($ \sim$ 6\arcsec) as that of the 250 $\micron$ image (Fig. \ref{fig:sedimage}). 
The \textit{Herschel} images show integrated emission from all the cloud components lying along the line-of-sight. 
To determine the density structure of the embedded cores, it is important to subtract the background emission which corresponds to the emission from the surrounding envelopes and large-scale cloud structure in all the images. 
The radial intensity profiles for our prestellar cores were approximated with a simple function of the form,
\begin{equation}
	I(R) = \frac{I_0}{1+(R/R_\mathrm{flat})^\alpha}+I_\mathrm{B}\; ,\label{equ:app1}
\end{equation}
where $I_0$ is the central peak intensity, $R_\mathrm{flat}$ is the radius of the flat region, $\alpha$ is the asymptotic power index, and $I_\mathrm{B}$ is the background level, which is a simple modification of the relation between $n(r)$ and $n_0$ to $I(R)$ and $I_0$ in the density profile of a starless core, $n(r) = n_0/[1+(r/r_\mathrm{flat})^\alpha]$ \citep{Tafalla:2002bn}. 
The original form of this function is meant to imply that the continuum emission asymptotically approaches to zero at a large radius. 
But in reality, the {\textit{Herschel}} images still have a considerable intensity at radii larger than the core radius, which corresponds to the background brightness. 
We accounted this background emission in our fitting procedure by including an additional term, $I_\mathrm{B}$, in Equ. \ref{equ:app1}. 
Using this function, we fitted a radial profile to the dust emission from each of the cores and determined the level of background emission that needs to be subtracted (Table \ref{tbl:dustfit}) from each of the images.

The observed dust intensity ($I_\nu$) can be simplified under the optically thin condition as 
\begin{equation}
	I_\nu = \mu_\mathrm{H_2} m_\mathrm{H} \kappa_\nu N_\mathrm{H_2} B_\nu (T_\mathrm{d})\; ,
\end{equation}
where $I_\nu$ is the intensity of dust emission for each pixel, $\mu_\mathrm{H_2}$ is the mean molecular weight per hydrogen molecule, $m_\mathrm{H}$ is the H-atom mass, $\kappa_\nu$ is the opacity,
\begin{equation}
	\kappa_\nu = 0.1 \left( \frac{\nu}{1000\;\mathrm{GHz}} \right)^\beta \;[\mathrm{cm^2\;g^{-1}}]
\end{equation}
($\beta$ is assumed to be 2), and $B_\nu (T)$ is the Planck function, 
\begin{equation}
	B_\nu (T_\mathrm{d}) = \frac{2 h \nu^3}{c^2} \frac{1}{\exp(h \nu / k T_\mathrm{d}) - 1}
\end{equation}
\citep{Kauffmann:2008jj}. 
The function was used to make the SED fit for dust emission from 160 to 500 \micron. 
The {\tt\string curve\_fit} used in the iteration was found to be extremely useful in extracting the best results on $T_\mathrm{d}$ and $N_\mathrm{H_2}$ which give the lowest $\chi^2$ fit value. 
The parameter space for $T_\mathrm{d}$ and $N_\mathrm{H_2}$ were searched for in the range of $5-20$ K and $10^{20}-10^{24}$ cm$^{-2}$, respectively. 
The location of $\chi^2$ minimum which gives the best-fit parameters was always well defined as a single localized position in $\chi^2$ distribution.

Figure \ref{fig:sedimage} shows the full data set and the results from the SED fit. 
We note that the dust temperature clearly shows a tendency to decrease toward the center of the core. 
The column density distribution of the \hmol\ gas is similar to the distribution of the dust continuum as a whole, but the position of the maximum point is slightly off because the position of the lowest temperature does not exactly coincide with the peak position of the dust continuum emission.

\subsection{$\mathrm{CS}$ and \nthp}\label{app:cdcal-mols}

The $N(\mathrm{N_2H^+})$, \nthp\ column density is calculated from our NRO observations by following the procedure described in the Appendix A of \citet{Caselli:2002eg}. 
Since \nthp\ $(1-0)$ line is optically thin, we use equation of total column density as
\begin{eqnarray} \label{eq:nmol}
	N & = & \frac{8\pi W}{A} \frac{\nu^3}{c^3} \frac{g_\mathrm{l}}{g_\mathrm{u}} \frac{1}{J_\nu (T_\mathrm{ex}) - J_\nu (T_\mathrm{bg})} \frac{1}{1 - \,\mathrm{\;exp} (-h\nu / k T_\mathrm{ex})} \nonumber \\ 
	& & \frac{Q}{g_\mathrm{l} \,\mathrm{\;exp} (-E_\mathrm{l} / k T_\mathrm{ex})} \; ,
\end{eqnarray}
where $W$ is an integrated intensity of the observing line, $\nu$ is the rest frequency of the line, respectively. 
$g_\mathrm{l}$ and $g_\mathrm{u}$ are the statistical weight of lower and upper levels, $J_\nu (T_\mathrm{ex})$ and $J_\nu (T_\mathrm{bg})$ are the equivalent Rayleigh-Jeans excitation and background temperatures, respectively. 
$T_\mathrm{ex}$ is the excitation temperature which is assumed to be the same for all rotational levels and was given as $T_\mathrm{d}$. 
$Q$ is the partition function and $E_\mathrm{l}$ is the energy level at the lower transition, given for \nthp\ and CS as linear molecules by, 
\begin{equation}
	Q = \displaystyle\sum_{J=0}^{\infty} (2J+1) \,\mathrm{\;exp} (-E_J / kT)\; ,
\end{equation}
and
\begin{equation}
	E_J = J (J+1) hB\; ,
\end{equation}
where $J$ is the rotational quantum number and $B$ is the rotational constant.

The $N(\mathrm{CS})$, CS column density, is determined by multiplying the \ctfs\ column density and the isotopic ratio of 22.7 \citep{Lucas:1998ue}. 
As \ctfs\ $(2-1)$ line is optically thin like \nthp, we therefore used the same equations mentioned above but with its corresponding parameters. 
The parameters used to calculate the molecular column densities are summarized in Table \ref{tbl:const}.

\bibliography{refs}
\end{document}